\documentstyle[aps,epsfig]{revtex}
\tightenlines
\begin{document}
\draft
\title{Search for Strange Quark Matter Produced in Relativistic Heavy 
Ion Collisions}
\maketitle
\begin{center}
\mbox{T.A. Armstrong                \unskip,$^{(9)}$}
\mbox{K.N. Barish                   \unskip,$^{(4,\ast)}$}
\mbox{S. Batsouli                  \unskip,$^{(14)}$}
\mbox{S.J. Bennett                  \unskip,$^{(13)}$}
\mbox{M. Bertaina                   \unskip,$^{(8,\dag)}$}
\mbox{A. Chikanian                  \unskip,$^{(14)}$}
\mbox{S.D. Coe                      \unskip,$^{(14,\ddag)}$}
\mbox{T.M. Cormier                  \unskip,$^{(13)}$}
\mbox{R.R. Davies                   \unskip,$^{(10,\S)}$}
\mbox{G. DeCataldo                  \unskip,$^{(1)}$}
\mbox{P. Dee                        \unskip,$^{(13)}$}
\mbox{G.E. Diebold                  \unskip,$^{(14)}$}
\mbox{C.B. Dover                    \unskip,$^{(2,\|)}$}
\mbox{P. Fachini                    \unskip,$^{(13)}$}
\mbox{B. Fadem                      \unskip,$^{(6)}$}
\mbox{L.E. Finch                    \unskip,$^{(14)}$}
\mbox{N.K. George                   \unskip,$^{(14)}$}
\mbox{N. Giglietto                  \unskip,$^{(1)}$}
\mbox{S.V. Greene                   \unskip,$^{(12)}$}
\mbox{P. Haridas                    \unskip,$^{(8,\P)}$}
\mbox{J.C. Hill                     \unskip,$^{(6)}$}
\mbox{A.S. Hirsch                   \unskip,$^{(10)}$}
\mbox{R.A. Hoversten                \unskip,$^{(6)}$}
\mbox{H.Z. Huang                    \unskip,$^{(3,\ast\ast)}$}
\mbox{H. Jaradat                    \unskip,$^{(13)}$}
\mbox{B. Kim                        \unskip,$^{(13)}$}
\mbox{B.S. Kumar                    \unskip,$^{(14,\dag\dag)}$}
\mbox{T. Lainis                     \unskip,$^{(11)}$}
\mbox{J.G. Lajoie                   \unskip,$^{(6,\S\S)}$}
\mbox{R.A. Lewis                    \unskip,$^{(9)}$}
\mbox{Q. Li                         \unskip,$^{(13)}$}
\mbox{B. Libby                      \unskip,$^{(6,\ddag\ddag)}$}
\mbox{R.D. Majka                    \unskip,$^{(14)}$}
\mbox{T.E. Miller                   \unskip,$^{(12)}$}
\mbox{M.G. Munhoz                   \unskip,$^{(13)}$}
\mbox{J.L. Nagle                    \unskip,$^{(5,\S\S)}$}
\mbox{I.A. Pless                    \unskip,$^{(8)}$}
\mbox{J.K. Pope                     \unskip,$^{(14,\|\|)}$}
\mbox{N.T. Porile                   \unskip,$^{(10)}$}
\mbox{C.A. Pruneau                  \unskip,$^{(13)}$}
\mbox{M.S.Z. Rabin                  \unskip,$^{(7)}$}
\mbox{J.D. Reid                     \unskip,$^{(9,\ast\ast\ast)}$}
\mbox{A. Rimai                      \unskip,$^{(10,\dag\dag\dag)}$}
\mbox{A. Rose                       \unskip,$^{(12)}$}
\mbox{F.S. Rotondo                  \unskip,$^{(14,\ddag\ddag\ddag)}$}
\mbox{J. Sandweiss                  \unskip,$^{(14)}$}
\mbox{R.P. Scharenberg              \unskip,$^{(10)}$}
\mbox{A.J. Slaughter                \unskip,$^{(14)}$}
\mbox{G.A. Smith                    \unskip,$^{(9)}$}
\mbox{P. Spinelli                   \unskip,$^{(1)}$}
\mbox{M.L. Tincknell                \unskip,$^{(10,\P\P)}$}
\mbox{W.S. Toothacker               \unskip,$^{(9)}$}
\mbox{G. Van Buren                  \unskip,$^{(8,\S\S\S)}$}
\mbox{W.K. Wilson                   \unskip,$^{(13)}$}
\mbox{F.K. Wohn                     \unskip,$^{(6)}$}
\mbox{E.J. Wolin                    \unskip,$^{(14,\P\P\P)}$}
\mbox{Z. Xu                         \unskip,$^{(14)}$}
\mbox{K. Zhao                       \unskip,$^{(13)}$}
\it
\centerline{(The E864 Collaboration)}
  $^{(1)}$ University of BARI/INFN, Bari, Italy \break
  $^{(2)}$ Brookhaven National Laboratory, Upton, New York 11973 \break
  $^{(3)}$ University of California at Los Angeles, Los Angeles,
California 90095 \break  
  $^{(4)}$ University of California at Riverside, Riverside, California 
92521 \break
  $^{(5)}$ Columbia University, New York, New York 10027 \break
  $^{(6)}$ Iowa State University, Ames, Iowa 50011 \break 
  $^{(7)}$ University of Massachusetts, Amherst, Massachusetts 01003
\break
  $^{(8)}$ Massachusetts Institute of Technology, Cambridge,
Massachusetts 02139 \break 
  $^{(9)}$ Pennsylvania State University, University Park, Pennsylvania 
16802 \break 
  $^{(10)}$ Purdue University, West Lafayette, Indiana 47907 \break 
  $^{(11)}$ United States Military Academy, West Point, New York 10996 
\break
  $^{(12)}$ Vanderbilt University, Nashville, Tennessee 37235 \break 
  $^{(13)}$ Wayne State University, Detroit, Michigan 48201 \break 
  $^{(14)}$ Yale University, New Haven, Connecticut 06520 \break
\end{center}


\date{\today} \maketitle

\begin{abstract}
We present the final results from Experiment 864 of a search for charged
and neutral strange quark matter produced in interactions of 11.5 GeV/c 
per nucleon $Au$ beams with $Pt$ or $Pb$ targets.  Searches were made
for strange quark matter with $A\geq5$.  Approximately 30 
billion 10\% most central collisions were sampled and no strangelet 
states with $A\leq100$ were observed.  We find 90\% confidence level
upper limits of approximately $10^{-8}$ per central collision for both 
charged and neutral strangelets.  These limits are for strangelets with 
proper lifetimes greater than 50 ns.  Also limits for $H^{0}$-d and
pineut production are given.  The above limits are compared with the 
predictions of various models.  The yields of light nuclei from 
coalescence are measured and a penalty factor for the addition of one 
nucleon to the coalescing nucleus is determined.  This is useful in 
gauging the significance of our upper limits and also in planning future
searches for strange quark matter.  
\end{abstract}
\vspace{0.2in}

\section{Introduction}

Color singlet states observed so far\footnote{Evidence for a
$q\overline{q}q\overline{q}$ state has been reported\cite{adams}}
consist of three quarks (baryons), three antiquarks (antibaryons) or 
quark-antiquark pairs (mesons).  These states are described by the 
Standard Model
which does not forbid the existence of color singlet states in a bag
containing an integer multiple of three quarks.  In such quark matter 
states all the quarks are free within the hadron's boundary and so 
are inherently different from nuclear states that are composed of a
conglomerate of A = 1 baryons.  Quark matter states composed of only up 
and down quarks are known to be less stable than normal nuclei of the 
same baryon number A and charge Z since nuclei do not decay into quark 
matter.  This is because of the relatively large
Fermi energy of two-flavor quark matter.  

Strange quark matter (SQM), composed of strange as well as up and down 
quarks, has several stabilizing factors that could result in quasistable
states.  The 
presence of strange quarks lowers the Fermi energy and the most stable
configurations for a given A would have roughly equal numbers of up,
down and strange quarks with charges of +2/3e, -1/3e and -1/3e,
respectively, therefore minimizing the surface and Coulomb energies.  
A major destabilizing factor is the large mass of the strange quark.
The above factors imply that the most stable varieties of strange quark 
matter should have a low value of Z/A and increase in stability with 
mass number.  The property of low Z/A provides the basis for current SQM
searches at heavy ion accelerators.  

\subsection{Theoretical predictions for strange quark matter}

Chin and Kerman \cite{chin} in 1979 predicted that SQM with $A\leq10$
might be metastable with halflife$\leq10^{-4}s$.  These predictions used
quantum chromodynamics (QCD) and the MIT bag model of hadrons \cite{bag}
to treat SQM quantitatively.  Subsequently, similar calculations with
the addition of shell effects were
carried out by Farhi and Jaffe \cite{farhi} and Gilson and Jaffe
\cite{jaffe}.  All theories contain the prediction that SQM systems
become more stable as A increases due to the small total charge of SQM
and bag model effects.  For sufficiently large A ($A\sim100$ to
$A\sim10000$, depending on the parameters assumed), SQM might be
absolutely stable \cite{witten}.  At the low-mass end Jaffe
\cite{dibaryon} proposed the existence of a neutral metastable dibaryon 
called the $H^0$ consisting of (uuddss) quarks.  Its
lifetime was estimated \cite{donog} to be less than $\sim2\times10^{-7}$.

It has been postulated that there may exist compact astrophysical
objects composed entirely of strange matter called strange stars.
Several astrophysical mechanisms are available to convert very large
stars to strange stars as discussed by Li \emph{et al.}\cite{li} and
references therein.  They also postulate that the millisecond pulsar SAX
J1808.4-3658 is a good candidate for a strange star. 

For smaller A, SQM may be metastable if strong decays are forbidden, but
could undergo weak decays with lifetimes in the range from 
$10^{-4}$ to $10^{-10}$ sec \cite{chin,berger,shaw}.  Effects of
\emph{Pauli blocking} may help to increase SQM lifetimes.  Such systems 
with $A\leq{100}$ which might be produced in relativistic heavy ion
collisions are commonly called \emph{strangelets} and are predicted to
be metastable for a wide range of SQM properties and bag model
parameters \cite{farhi,jaffe,shaw}.  Due to the lack of theoretical 
constraints on bag model parameters and difficulties in calculating
color magnetic interactions and finite size effects \cite{schaff,madsen}
experiments are necessary to help answer the question of the stability
of strangelets if indeed they do exist.  
        
Relativistic heavy ion collisions provide a promising mechanism for
producing strangelets in the laboratory due to the high baryon densities
and the large number of strange quarks achieved in a small volume during
these collisions.  Several classes of models have been generated to 
describe strangelet production in nucleus-nucleus collisions.  They can 
be classified into two categories, namely strangelet production by 
coalescence or strangelet production following quark-gluon plasma (QGP)
production.

In \emph{coalescence models}
\cite{baltz} a number of A=1 particles are produced in the collision
that in turn fuse to form a strangelet.  \emph{Thermal models} further 
assume that thermal and chemical equilibrium are achieved prior to the 
production of the final particles \cite{pbm}.  Coalescence and thermal 
models usually predict lower strangelet cross sections than models that 
postulate a collision in which a QGP state is formed. 

It might be possible to produce a phase transition to a QGP in these 
collisions.  Under these conditions the hot quark matter might cool into
a metastable state of cold SQM resulting in a strangelet.  Models have 
been produced to examine production of
strangelets following QGP formation.  Kapusta \emph{et al.} estimate
that at AGS energies there could be rare events in which a droplet of
QGP is nucleated converting most of the superheated matter to plasma 
\cite{kapusta}.  They calculate the probability that
thermal fluctations in a superheated hadronic gas will produce a thermal
droplet and that the droplet will be large enough to overcome its
surface free energy and grow.  They estimate this to occur in between 
0.1\% and 1\% of central (small impact parameter) $Au$ + $Au$ collisions
at AGS energies.  

Greiner \emph{et al.} suggests that once a QGP droplet is formed, for a 
wide range of QGP properties, almost every QGP state evolves into a 
strangelet by means of the strangeness distillation mechanism providing 
strangelets are metastable \cite{greiner}.  The droplet cools by
emitting mesons but the $\overline{s}$ quarks preferentially joins with 
u and d quarks to form K mesons in
the baryon rich plasma formed at AGS energies.  This leaves the QGP
enriched in strangeness relative to anti-strangeness leading to the
formation of a strangelet during the hadronization process.  This
process favors the formation of the more stable large strangelets since
the QGP would lose energy by meson emission possibly resulting in a
strangelet of approximately the same A as the QGP droplet.  It is thus
important to carry out experiments that are sensitive to a large mass
range.  

Other estimates of strangelet production by distillation from the QGP
were carried out by Liu and Shaw \cite{liu} and Crawford
\cite{crawford}.  They predict a wide range of production levels.
Strangelet production could prehaps be as high as $10^{-4}$ to $10^{-3}$
per central $Au + Au$ collision at AGS energies.  Based on a recent
calculation, Schaffner-Bielich \emph{et al.} have suggested that at low
masses negative strangelets are more likely to be formed than positive 
strangelets \cite{schbie}.  They used the MIT bag model with shell mode 
filling for various bag parameters.  Their model also predicts a number 
of neutral strangelet states to be metastable.  

In this experiment we have searched for positive, negative and neutral 
strangelets with masses up to A=100.  We note that our apparatus would 
detect strangelets of $A\geq100$ if they were produced.  However, the 
production probability with coalescence would certainly vanish at baryon
numbers of the order of 100 or more.  The strangeness distillation model
would produce strangelets which at the extreme would be less than the 
total numner of baryons in the collision (197+208 for a $Au-Pb$ 
collision).  Furthermore, due to saturation in the response of the E864 
calorimeter, all masses higher than A=100 would be detected as having 
mass very close to A=100.  For these reasons we show our yields as
functions of A up to A=100.

\subsection{Previous searches for strange quark matter}

Searches for in situ SQM have been made on terrestrial matter 
\cite{hemmick}, cosmic rays and astrophysical objects \cite{alcock}.  
These searches resulted in extremely low limits for strangelets in 
terrestrial matter.  These rates are less than predicted by Big Bang 
models of strangelet production in the early universe and so would argue
against the existence of completely stable strangelets.  This conclusion,
however, is somewhat ambiguous due to the uncertainties in the models 
themselves, the uncertainty in estimating strangelet survival
probabilities and possible geophysical processes which could ``distill''
the terrestrial strangelets into unaccessible regions.

With the advent of relativistic heavy ion beams at the AGS and SPS 
accelerators it is possible to search for metastable strangelets in the 
reaction products from central collisions where a large number of
strange quarks are produced.  Searches for strangelets have been carried
out using relativistic heavy ion beams from the AGS and SPS
accelerators.  Early searches which used $Si$ + $Cu$ \cite{barette} and 
$Si$ + $Au$ \cite{aoki} reactions at the AGS accelerator and $S$ + $W$ 
\cite{borer} reactions at the SPS accelerator yielded null results.  
Later
experiments at the AGS accelerator using $Au$ beams \cite{beavis} and at
the SPS accelerator using $Pb$ beams \cite{appel} also yielded null 
results despite the increased production potential of these heavier 
beams.  To date no experiment has published results indicating a clear
positive signal for strangelets so all have set production upper limits.

The searches for strangelets discussed above using relativistic heavy
ion collisions were sensitive to proper lifetimes down to about 50 ns.
All of these experiments except the one carried out by E814
\cite{barette} used
focusing spectrometers which, for a given magnetic field setting, have 
good acceptance only for a fixed momentum and charge of the produced
particle.  Therefore, the production limits obtained in these
experiments are strongly dependent upon the production model assumed for
high mass particles such as strangelets.  

Recent searches for the $H^0$ dibaryon have been carried out.  If the
$H^0$ decays weakly by $\Delta S=+1$ then the expected lifetime is
similar to that of the $\Lambda$, therefore most searches look for decay
processes with $\Lambda$ type lifetimes.  However, if the $H^0$ is very
tightly bound, then only $\Delta S=+2$ decays are allowed and a
resulting lifetime of the order of 50 ns is possible.  Using the
1.8 GeV/c $K^-$ beam from the AGS, Stotzer \emph{et al.}\cite{stotzer}
in E836 searched for the $H^0$ at a mass range from 50 to 380
MeV/$c^{2}$ below the $\Lambda\Lambda$ threshold.  E810 and E888 have
also carried out searches for the $H^{0}$ dibaryon\cite{long,belz} at
the AGS by searching for it's decay modes but conclusive evidence for 
it's existence has not been obtained.  The neutral beam produced by 
800 GeV/c protons on a $BeO$ target was analyzed by the Fermilab KTeV
collaboration\cite{alavi} to search for the $H^0$.  No events consistent
with interpretation as an $H^0$ were observed.  A complete summary
of searches carried out for various forms of strange quark matter is 
given in a review article by Klingenberg \cite{kling}. 

The primary goal of the E864 experiment was to search for charged and 
neutral strangelets with lifetimes $\geq5\times10^{-8}$ seconds, baryon 
number A from 6 up to 100 
and charge to mass ratios lower than most normal nuclei.  These 
characteristics suggested a strategy of looking for mid-rapidity,
massive objects with an unusual Z/A ratio in an apparatus with a high 
rate capability and redundancy for background rejection.  The E864 
apparatus implemented this strategy as described in the next section.  
E864 results from earlier data sets with smaller statistics than the 
results shown here have been published for charged strangelets in a 
series of papers by Armstrong \emph{et al.} \cite{strange} as well as
for neutral strangelets \cite{prun}.  In this paper we give the final
limits for both charged and neutral strangelets from the E864
experiment.

\section{The E864 Experiment}

\subsection{General design of experiment}

The E864 experiment is an open geometry, two dipole magnetic
spectrometer designed to search for strangelets in $Au$ + $Pt,Pb$ 
collisions at 11.5 GeV/c per nucleon.  The experiment is described in 
detail in Ref~\cite{e864_nimpap}.  The open geometry with only dipole 
magnets causes the experiment to be less sensitive to the shape of a 
particle's production differential cross section.  Due 
to the nature of the design of the rare particle search, the
spectrometer is also well suited for detecting nuclear isotopes and 
hypernuclei produced by coalescence following central collisions.  

The spectrometer identifies particles via their mass M and charge Z.  In
order to conduct this search, E864 has a large geometric acceptance 
(5 msr) and operates at a high data rate.  The emphasis is on the 
measurement of particles near the center-of-mass rapidity since it leads
to an efficient search with minimal production model dependence.  A 
diagram of the spectrometer is shown in Figure \ref{fig:e864}.  The main
components are beam defining counters, a target system (usually $Pb$ or 
$Pt$), a multiplicity counter for triggering 
on the centrality of the event, two analysis dipole magnets, three 
stations of hodoscopes for Time-of-Flight (TOF) and tracking, two 
stations of straw tubes for tracking, a hadronic calorimeter and a large
vacuum tank not shown in this figure.  The experiment also utilizes a
high speed data acquisition system (2.7 Mbytes per second) and a
flexible second level trigger (Late Energy Trigger) based on TOF and 
energy as measured by the calorimeter.  

\subsection{Experimental details}

\subsubsection{Target area}

The E864 spectrometer receives a fully stripped $Au$ beam with a
momentum of 11.5 GeV/c per nucleon.  The ions are incident on a $Pb$ or
$Pt$ target of thickness between 5\% and 60\% of a $Au$ 
interaction length.  The nucleon-nucleon center of mass energy is 
4.6 GeV and its rapidity $y$ is 1.6.  The experimental layout in the 
target area consists of quartz Cherenkov beam counters (MITCH) and beam
defining counters and a scintillator 
multiplicity counter to select events with the desired centrality.  
Thin quartz plates are used for all counters traversed by the beam to 
minimize the number of interactions in the counters.  The beam counters 
measure the incident beam flux and provide the start time
for the hodoscope and calorimeter TDCs \cite{beam_nim}.  

The multiplicity counter consists of a four quadrant annulus placed 
around the beam pipe 13 cm downstream of the target.  It subtends an 
angular range of $16.6^o$ to $45.0^o$.  The  
total signal measured with this counter is proportional to the
centrality of the collision and is used to trigger on the centrality of
the events.  Most data is taken with a threshold to accept the 10\% of
events with the largest multiplicity counter signals.  

\subsubsection{Tracking systems}

\begin{figure}
\epsfig{bbllx=72pt,bblly=260pt,bburx=562pt,bbury=594pt,file=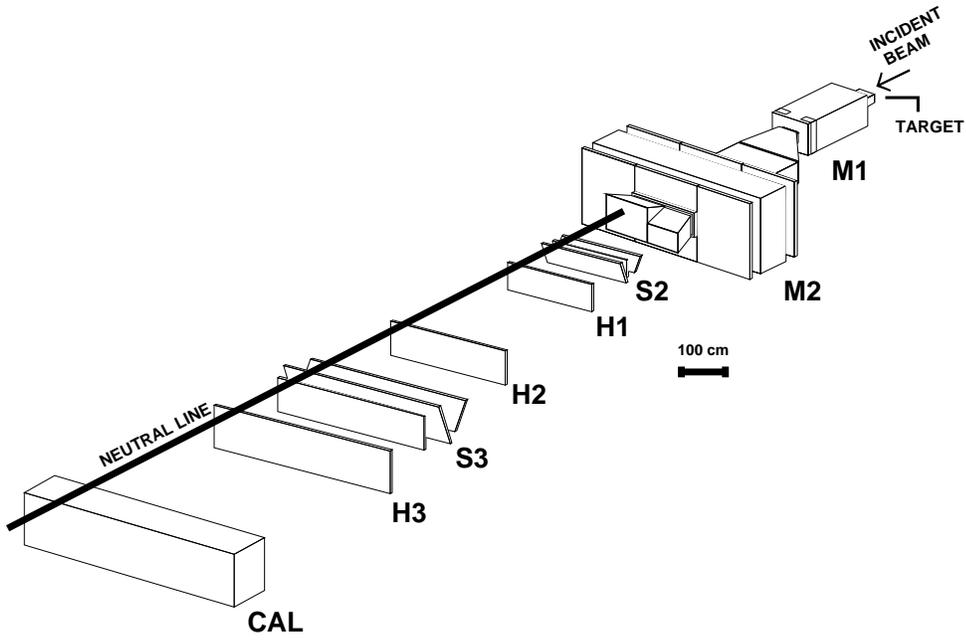,
width=5.0in}
\caption{Perspective view of the E864 spectrometer.  The vacuum tank is 
not shown.}
\label{fig:e864}
\end{figure}

The heart of the spectrometer tracking is the scintillator TOF 
hodoscope system which consists of three stations,  H1, H2 and H3, 
whose locations are shown in Figure \ref{fig:e864}.
The hodoscopes provide redundant measurements of a
particle's TOF as well as its position.  Requiring that a good space
track also have consistent velocities as measured at each of the 
hodoscopes significantly improves the background rejection.  The 
hodoscopes also give three independent charge measurements via the pulse
height information from energy loss ($dE/dx$) in the scintillator.  The 
hodoscopes provide $x-y-z$ space points.  The time resolution for H1 and
H2 is $140\pm10$ ps and for H3 is $160\pm10$ ps.  The measured 
efficiencies of the hodoscopes range from 97.7\% for H1 to 98.9\% for H2
and H3.

In order to improve the spatial resolution for tracked particles, the 
spectrometer has two stations of straw tubes, referred to as S2
and S3 in Figure~\ref{fig:e864}.  A complete description can be found in
Reference\cite{straw_nim}.  Each station consists of three 
sub-planes ($x,u,v$) each consisting of two layers.  The straws of the 
x plane are mounted vertically and the straws 
of the u and v planes are mounted at $\pm$20$^o$ relative to vertical,
respectively.  Each sub-plane consists of two staggered layers of straw 
tubes 4 mm in diameter.  The planes are rotated around the vertical
axis at approximately 6 degrees with respect to the beam line so that
most particles are incident perpendicular to the planes.  

A straw tube chamber S1 was placed inside the vacuum tank between M1 and
M2.  The chamber was designed to improve the tracking by providing a
track measurement between the magnets.  However, due to a problem with 
discharges associated with the high voltage connections, the chamber did
not work well enough to be useful.  Due to the exigencies of the 
experimental run, the chamber was left in place and contributed to 
background scattering processes.  The analysis which we carried out is 
correct but would have given a slightly more sensitive result if S1 had 
been removed.

\subsubsection{The calorimeter}

The final element of the spectrometer is a ``spaghetti'' design
hadronic calorimeter located at the end of the E864 beamline as shown in
Figure~\ref{fig:e864}.  Its purpose is to provide a second independent 
mass measurement for charged particles and to identify neutral particles
based on $\beta$ and the deposited energy.  The tower construction is 
based on a design first tested by the SPACAL collaboration\cite{spacal}.
The spaghetti design allows a close packed geometry and virtually 
eliminates gaps or dead regions in the detector fiducial volume.  This
results in very good energy and time resolution.  In addition, the 
detector response is quite uniform and nearly independent of the 
position of the particle.  Details of the construction and performance 
of the calorimeter have been published\cite{calo_nim}.

The calorimeter consists of 
$58~(\mathrm{horizontal}) \times 13~(\mathrm{vertical})$ towers. 
The whole assembly is rotated $3.3^o$
with respect to the beam direction.  The dimensions of each tower is 
$10~\rm{cm} \times 10~\rm{cm} \times 117~\rm{cm}$.  The tower width is 
smaller than the typical transverse size of a hadronic shower thus 
allowing for transverse shower profile information.  The time and 
energy resolution of the calorimeter are excellent.  The resolution 
for showers in a $5\times5$ array is given by 

\begin{equation}
\frac{\sigma(E)}{E}=(3.5\pm0.5)\% + \frac{(34.4\pm0.8)\%}{\sqrt{E(GeV)}}
\label{eq:calres}
\end{equation}

and the hadronic time resolution is better than 400 psec. 

\subsubsection{Data acquisition and trigger}

The Data Acquisition System (DAQ) is designed to
record 4000 events per AGS spill and typically 1800 events per spill
are recorded.  Signals from the counters 
and triggers are sent into digitizers in FASTBUS or CAMAC.  Event data 
are sent to memory buffers residing in VME which are capable of 
buffering an entire spill's worth of data.  The event fragments in each 
buffer are assembled in event builder modules and transferred to eight 
Exabyte 8mm tape drives.  More details and specifications are given 
in \cite{john_thesis,IEEE_daq}. 

The first level of the E864 two-level trigger selects events
where a good beam particle had the desired centrality.  The second level
selects events based on time and energy measurements in the calorimeter 
and is called the Late Energy Trigger (LET).  The level 1 trigger 
requires that the beam counter signal is consistent with a single Au 
ion with no hits in either of the veto counters and that no beam
particle is within a time window before and after the selected hit. 
The trigger could also be set to exclude events below a given
multiplicity.

\begin{figure}
\epsfig{bbllx=0pt,bblly=126pt,bburx=540pt,bbury=666pt,%
file=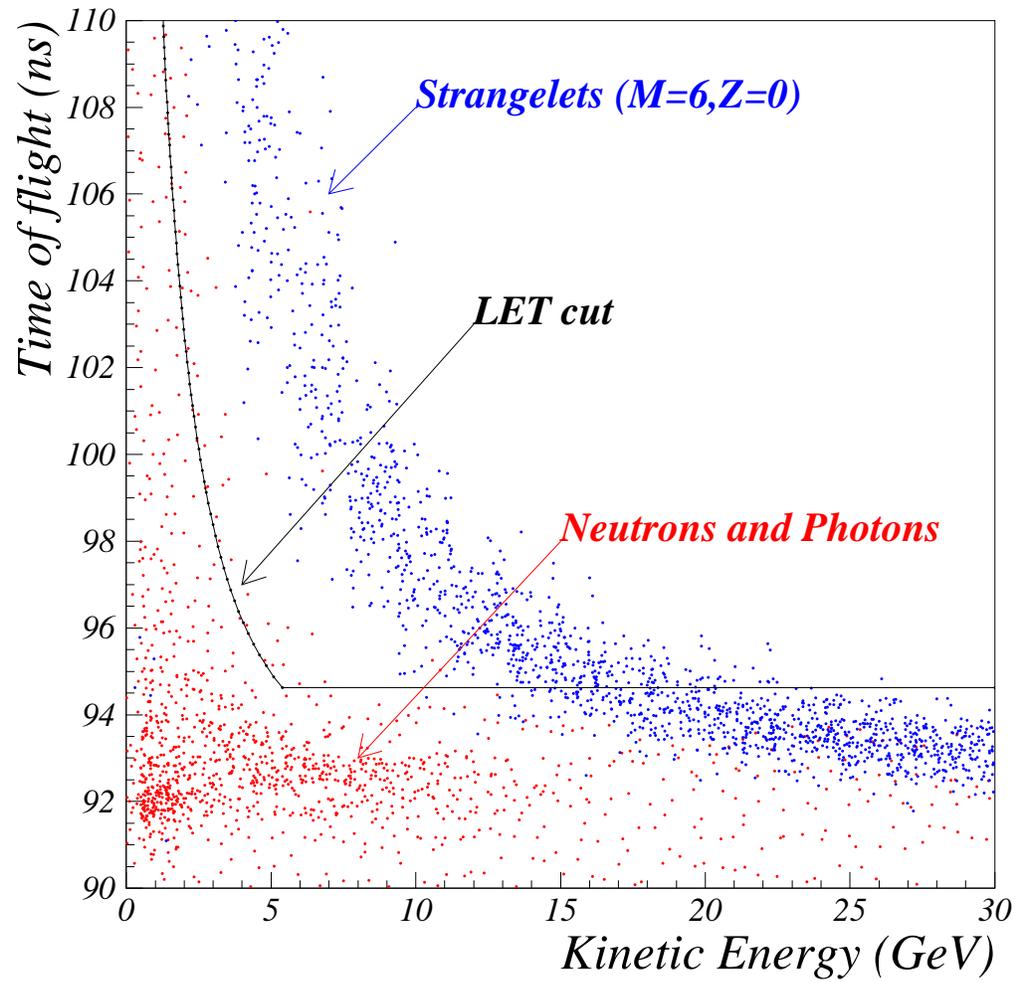,width=5.5in,clip=}
\caption{Simulation of the distribution of mass 6 uncharged strangelets,
protons and neutrons in time versus energy space.  The effects of
detector resolution have been included.  The solid curve illustrates a 
typical cut using the Late Energy Trigger (LET).}
\label{fig:let_perfect}
\end{figure}

The level 1 trigger provides sufficient rejection to study inclusive 
spectra of protons and kaons but a level 2 trigger is needed to obtain
the sensitivity required for the strangelet searches.  Since the 
calorimeter measures both energy and time in each tower, that 
information is used to determine the mass of the particle.  As an 
example Figure~\ref{fig:let_perfect} shows a simulation of the 
distribution in TOF versus energy for mass 6 uncharged strangelets 
compared to that for protons and neutrons with a curve to illustrate a 
typical cut with the effects of detector resolution included in the 
simulation.  The implementation of the LET is described in detail in 
Reference~\cite{let_nimpap}.

The LET system digitizes the energy and time signals from the
calorimeter providing indices into a programmable lookup table.  The 
output of the lookup tables is ORed to form an accept or 
reject.  The lookup table is generated in terms of energy
and TOF from Monte Carlo and data.  A typical trigger table efficiency 
is 85\% for a mass 5 GeV/$c^{2}$, charge +1 strangelet
in the rapidity range $1.6\pm 0.5$ at the +1.5 T field setting and
increasing to almost 100\% for higher masses.  The corresponding 
rejection factor for the above example is 80, giving enhancements 
(defined as rejection times efficiency) of about 68.

\subsubsection{Monte Carlo simulations and acceptance}

\begin{figure}
\psfig{file=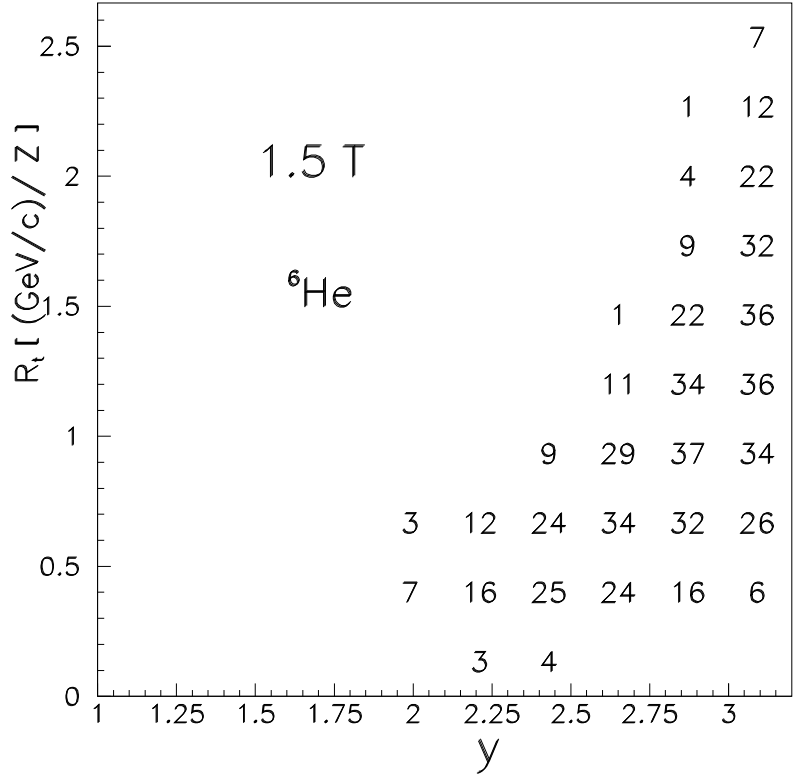,width=5.0in}
\vspace{-1.0in}
\caption{Acceptance of the spectrometer in transverse rigidity versus y
for $^{6}He$ for a 1.5 T field.  The acceptances are given in percent.}
\label{fig:he-6}
\end{figure}

   Extensive use of a GEANT3\cite{geant3} based Monte Carlo of the 
apparatus was made both in designing the shielding and detector as 
well as determining acceptances and efficiencies for the physics
results.  In the analysis stage, acceptances and efficiencies are 
obtained by tracking single particles with various production models 
through the GEANT3 model.  The acceptance of the spectrometer for
neutral particles is determined by the physical apertures of the 
collimators and magnets.  For charged particles 
with momentum $p$ and transverse momentum $p_t$, the acceptance in 
rigidity R=$p$/Z and transverse rigidity $\mathrm{R}_t$=$p_t$/Z
is constrained by the field of the magnets as well as these apertures.
For high positive fields the pions and protons are largely swept out of
the spectrometer acceptance.  This is a desirable feature when
searching for rare high mass objects such as strangelets.  There are 
regions in $y$ and $p_t$ with acceptance  
for the same particle species in different field settings.  This
provides an important check on the systematics.  In 
Figure~\ref{fig:he-6} the acceptance is shown for a heavy species, 
namely $^{6}He$, as a function of transverse rigidity and rapidity at a 
magnetic field of 1.5 T.  

\section{Data Analysis for Charged Strangelets}

\subsection{Determination of particle mass and charge}
   
The reconstruction of charged particle tracks uses information from 
the hodoscopes and the straw tubes.  The tracking algorithm begins by 
using the three-dimensional space hits in the hodoscopes to define 
straight line tracks downstream of the magnets in the $x-z$ and $y-z$ 
planes.  Consistent hits in the straw tubes are then attached to the 
track and the tracks are refit. Next the rigidities ($p$/Z) and path 
lengths of the tracks are determined from a lookup table whose inputs
are the $x-z$ and $y-z$ slopes of the tracks downstream which are assumed
to come from the target.  The lookup table is determined from a Monte 
Carlo simulation of the apparatus which includes a model of the magnetic
fields.  The table is only a few thousand entries in length due
to a sophisticated multi-dimensional interpolation\cite{alexi}. The 
method is very fast and has an intrinsic resolution of better than
0.1\%.  Next the path length versus TOF at the target and each of the 
hodoscopes is fit to determine the velocity of the tracked particle. 
Using the rigidity and the velocity, the track is refit using a
full multiple scattering correlation matrix.  The complete formalism is 
given in Appendix A of Ref~\cite{john_thesis}.  There are four fits: 
$x-z$, $y-z$, time vs pathlength and $y$-pathlength.  The track quality 
is evaluated by considering the chi-squareds of these fits.  Tracks with
a large chi-squared have a high probability of being associated with 
background processes.

Each of the x, u and v straw tube planes consists of two layers.  In
track reconstruction, each set of 2 layers is considered as one logical 
plane called a doublet.  The hits are combined into clusters which are 
groups of contiguous hits in the doublet.  Thus most clusters consist of
two hits, one from each plane.  The efficiency of each plane is
measured by leaving the plane of interest out of the track fits and then
checking if there is a hit in that plane consistent with the track. The
doublet efficiency, defined as the efficiency for having at least one
hit in the doublet, is typically 95\%-98\%.

One of the most important aspects of the spectrometer is its ability
to track particles in time as well as in space.  The arrival time of a 
charged particle relative to the arrival of the beam particle at the 
target is determined independently in the three hodoscope walls as well
as in the calorimeter.  The time of flight for a hit in a hodoscope slat
is given by 

\begin{equation}
T = \frac{1}{2}(TDC_{top}+TDC_{bot}) - T_{beam} +T_0
\label{eq:tof}
\end{equation}

where $TDC_{top}$ and $TDC_{bot}$ are the raw TDC values, corrected for
slewing and for differences in cable lengths and any time dependent
variations in the PMTs, cables and TDCs and converted to nanoseconds
using the time calibration of the TDCs.  $T_{beam}$ is the mean time for
the beam counter and is subtracted off event by event in order to remove
variations in the experimental gate.  $T_0$ is an offset which turns the
number into a true time of flight.  It is determined originally from MC
calculations and then fine tuned using tracked particles to calculate 
$\beta$=v/c from the measured momentum and assumed mass of the track.  
Note than an error in the magnetic field can be compensated for in
this constant if only one particle species is considered.  This is 
avoided by using particles of different species.  The $\beta$ of a 
particle is determined from a least square fit of path length from the 
target to the hodoscope planes versus TOF.  

\begin{figure}
\epsfig{file=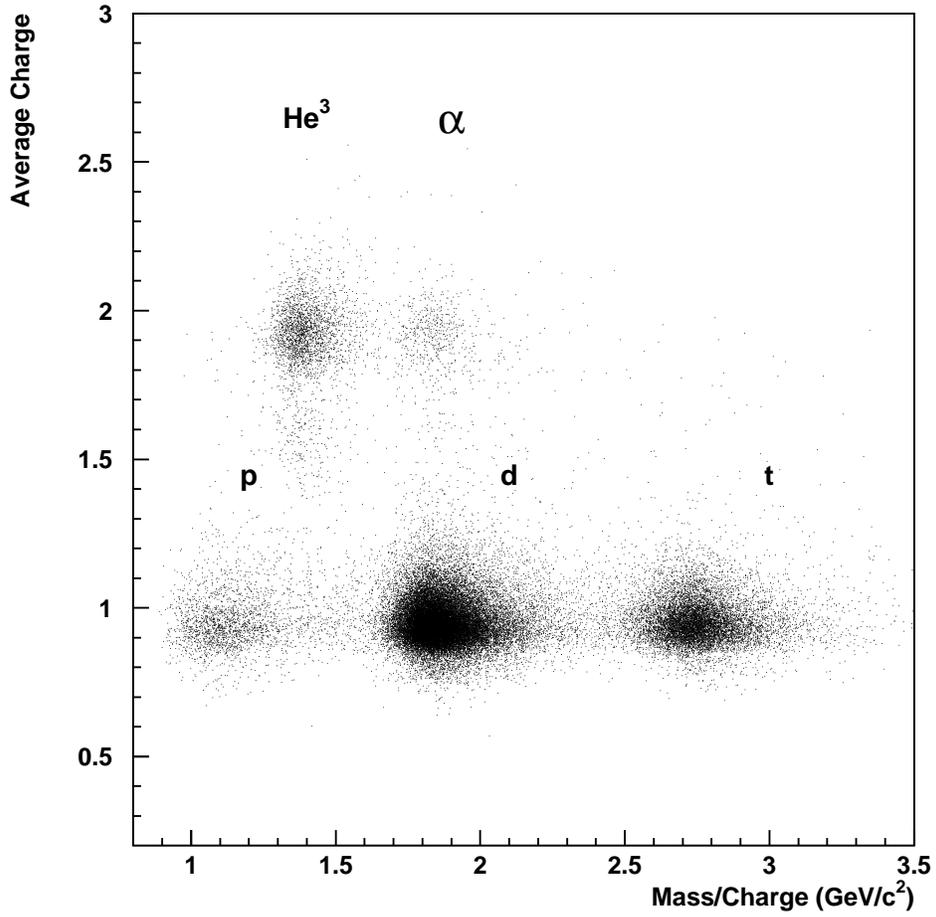,width=6.0in}
\vspace{-0.5in}
\caption{Particle identification using the charged particle tracking
system. The average of the charges as determined by the three hodoscopes
is plotted versus the mass/charge for positively charged tracks.  Clear 
peaks for abundant particle species are apparent. A $\beta$ cut of 
$<0.985$ has been applied.  The data are from a +1.5 T field run with the
LET trigger set to enhance higher mass objects.}
\label{fig:mass_charge}
\end{figure}

The charge $Z$ of a track is determined independently in each hodoscope 
wall using the the geometric mean of the measurements by the ADCs at the
top and bottom of the slat.  The geometric mean is used because it does 
not depend on the vertical position of the hit in the slat.  
Specifically, 

\begin{equation}
Z^2 = \sqrt{G_{top}(ADC_{top}-PED_{top})G_{bot}(ADC_{bot}-PED_{bot}) }
\label{eq:charge}
\end{equation}

where $G_i$, $ADC_i$, and $PED_i$ are the gain, ADC value and pedestal
for the top and bottom signals, respectively.  The pedestals are 
determined from ``empty'' events taken randomly throughout the spill.  
The gains are normalized for every slat by using tracked particles.
Typical efficiencies for the cuts used to isolate charge $\pm 1$ 
particles are $\approx 97\%$ per plane, or 91\% since all three planes 
are used.  Charge 2 efficiencies are somewhat lower, $\approx 93\%$ per 
plane or 80\% total.  The ability of the spectrometer to identify 
particles via 
their mass and charge is demonstrated in Figure~\ref{fig:mass_charge}
from the the 1995 data at +1.5 T.  The only cuts that are applied 
are chi-squared cuts on the tracks and $\beta\ <\ .985$.  Note that this
data is taken with the LET set to enhance higher mass particles. The 
various species are well separated and there is little background.

\begin{figure}
\epsfig{bbllx=0pt,bblly=125pt,bburx=585pt,bbury=715pt,%
file=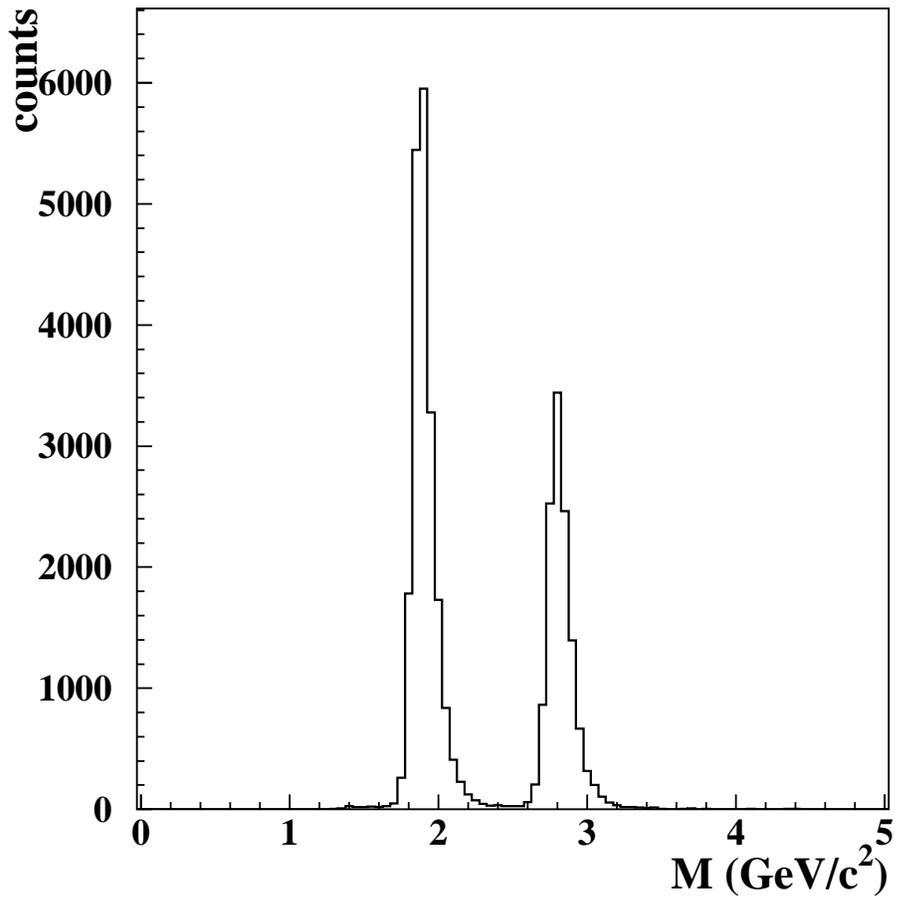,width=5.5in,clip=}
\caption{Single particle mass distributions for +1 charged particles at 
a 1.5 T field setting.  A $\beta$ cut of 0.972 was applied.  The data
are from the 1996/7 run using the LET.}
\label{fig:mass_15}
\end{figure}

\begin{figure}
\psfig{file=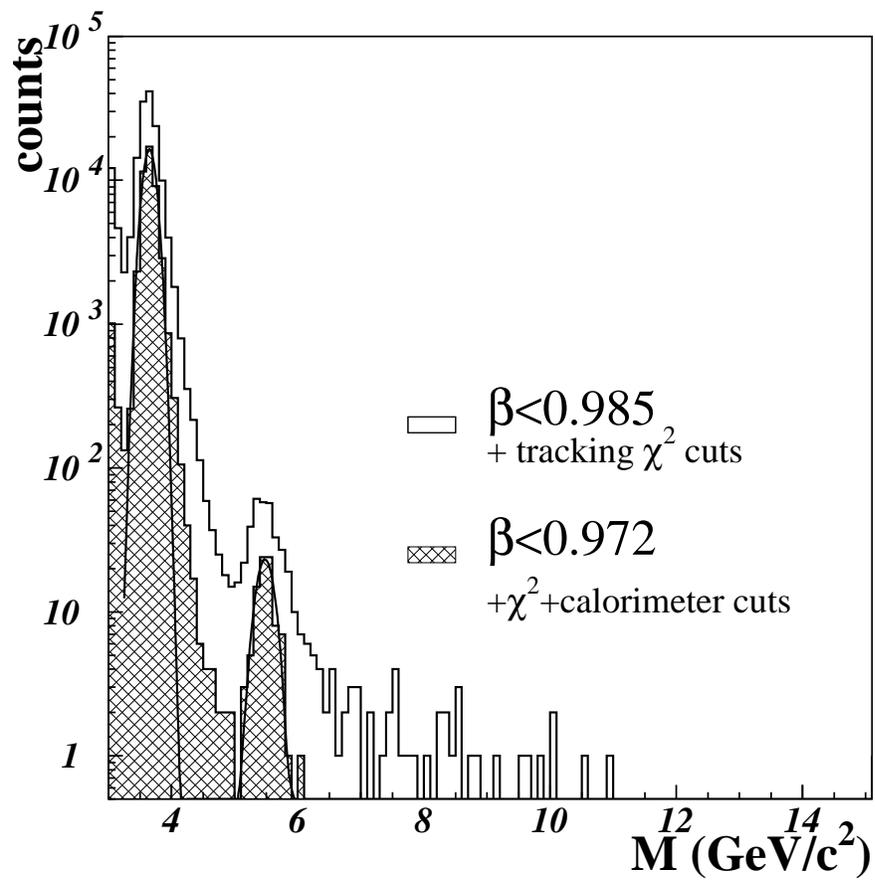,width=5.5in}
\vspace{-0.5in}
\caption{Single particle mass distribution for +2 charged particles at a
1.5 T field setting.  The two curves demonstrate the effect of
tightening the $\beta$ cut and adding calorimeter cuts.}  
\label{fig:zxb2}
\end{figure}

The single particle mass spectra at the 1.5 T field setting is shown in 
Figures \ref{fig:mass_15} and \ref{fig:zxb2} for charge +1 and +2 
particles, respectively.  The mass resolutions are on the order of
3-5\% and the peaks are very clean with minimal background.  Also the 
same species are accepted, although with different efficiencies, in
more than one field setting.  Requiring that results agree from one
field setting to the next provides an important check of
systematics, particularly
for invariant cross sections as a function of rapidity and $p_t$.  In 
Figure \ref{fig:zxb2} peaks from $^{4}He$ and $^{6}He$ are clearly seen.
Figure \ref{fig:zxb2} also demonstrates the benefit of calorimeter
cuts in eliminating charged particle background generated by charge 
exchange scattering of neutrons. 

\begin{figure}
\epsfig{bbllx=0pt,bblly=135pt,bburx=522pt,bbury=720pt,file=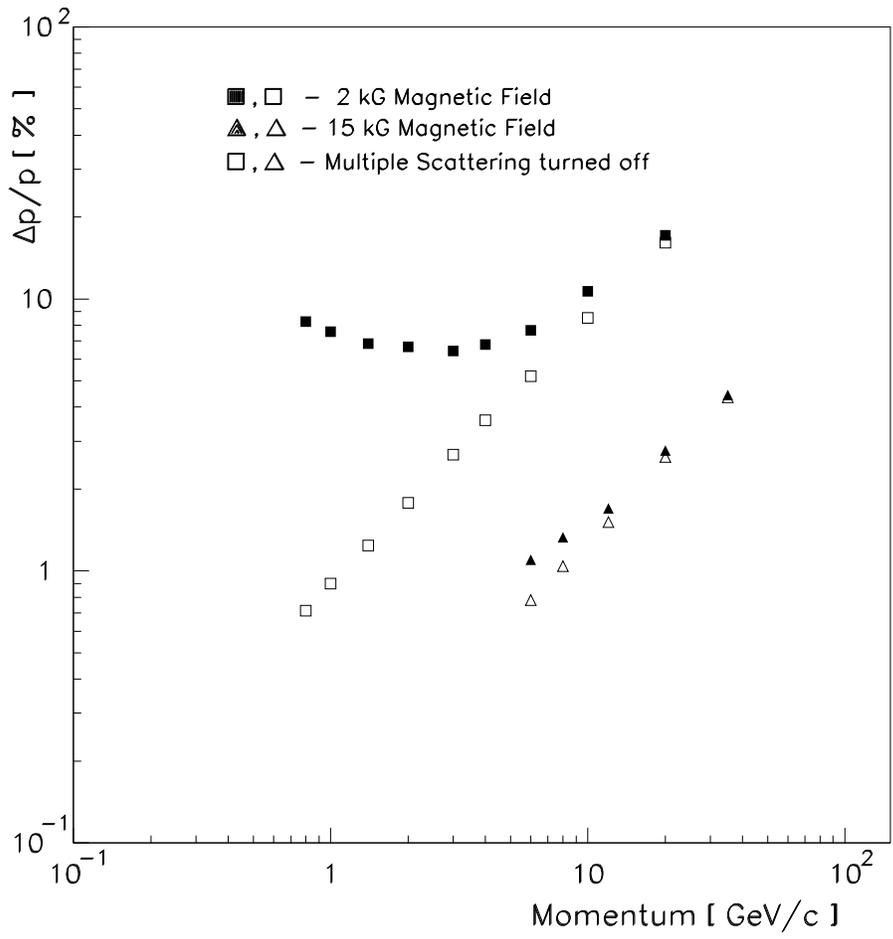,
width=5.0in}
\caption{Momentum resolution as a function of p for 0.2 T and 1.5 T 
magnetic fields.}
\label{fig:p_res}
\end{figure}

The mass is given by 

\begin{equation}
m = \frac{R \times Z}{\gamma\beta}
\label{eq:mass}
\end{equation}

where R is the particles rigidity.  The mass resolution is dependent on
the resolution of both $\beta$ (from TOF) and momentum.   The momentum 
resolution is given by

\begin{equation}
\frac{\sigma_{p}^{2}}{p^{2}} \approx \frac{\sigma_{B}^{2}}{B^{2}} + 
\frac{2\sigma_{\theta}}{\theta^{2}} 
\label{eq:massres}
\end{equation}

where $B$ is the magnetic field and $\theta$ is the angle of the track
in the bend plane as measured by the downstream tracking chambers.  The 
magnetic fields are known to $\approx \pm 1\%$.  The resolution in 
$\theta$ is determined by the 
multiple scattering (proportional to $1/p$) and the resolution of the 
straw tubes.  $\theta$ itself is proportional to the total field times 
$Z/p$.  Figure~\ref{fig:p_res} demonstrates these effects.  It gives the
momentum resolution $\sigma_p$ as a function of momentum $p$ for 0.2T
and 1.5 T fields for a charge 1 particle.  The open symbols are the 
distributions when multiple scattering is turned off in the simulation.  

\subsection{Background}

    The principal backgrounds in E864 are expected to be those which 
produce real tracks with the same directions and velocities as the 
tracks of interest.  Sources of such tracks are:
\begin{enumerate}
\item Overlapping events caused by two beam particles within 
the event time window of the detector, $\approx$50 ns.  Both interact in
the target or the later one interacts upstream of the target.  The
timing is set by the first one, so tracks from the second interaction
will be late, leading to an incorrect $\beta$ that is too small.
\item Charged tracks that originate downstream of the target, many of
which are created in interactions by neutrons generated in the target.
The track will be properly reconstructed downstream of the magnets, but 
when the track is extrapolated back to the target, the momentum will be
larger than it should be.  Sources of such tracks are secondary 
interactions the most troublesome of which are charge exchange of 
neutrons into protons in the vacuum chamber exit window just downstream 
of M2, in the air before S2 or in the first mono-layer of S2.  An
additional source of background was scattering of particles by the S1
straw tube array.
\end{enumerate}

The first class of backgrounds is minimized with veto counters and
the detection of multiple beam tracks in the trigger counters.  The 
second class of background is minimized by requiring that the
momentum as measured by the tracking chambers agree with the energy as 
measured in the calorimeter.  

\section{Data Analysis for Neutral Strangelets}

We report here the results of a search for neutral strangelets in 
relativistic heavy ion collisions.  The first information available on 
neutral strangelet limits was published\cite{prun} based on an earlier 
E864 data set with smaller statistics.  Background 
problems associated with searches for neutral particles are more severe.
In addition to all the background associated with charged particle 
searches, backgrounds are present due to the inability to 
track neutral particles.  The search for neutral strangelets capitalizes
on the excellent performance of the E864 spectrometer for the study of 
both charged and neutral hadrons.  The key element making the search for 
neutral strangelets possible is the hadronic calorimeter at the
downstream end of the E864 tracking system.  

\subsection{Search procedure}

The search for neutral strangelets is performed in three steps.  The
first step is to search for interesting hits in the calorimeter.  The
second step is to eliminate all hits corresponding to charged particles
reaching the calorimeter.  The final step is to eliminate clusters with
energy contamination from overlapping showers or products from late
interactions in the target.  In the first step the entire fiducial
volume for
the hadronic calorimeter is searched for each event to identify
particle hits that could represent an interesting object.  Hits that
represent a local maximum in energy and that also fired the LET are
selected for further analysis.  The particle energy is determined from
a sum, $E_{3x3}$, of $3x3$ towers surrounding the peak tower.  This 
corresponds on the average to 90\% of the total deposited energy.  

In the second step tracks are reconstructed using the
three planes of the hodoscopes and the straw tube chambers S2 and S3
in order to eliminate hits in the calorimeter from charged particles.
It is necessary to have a high efficiency for track reconstruction but
at the same time avoid false rejection of neutral particles due to ghost
tracks, therefore two different procedures are used for track 
reconstruction.  For neutral strangelet candidates with baryonic mass
less than 30 much contamination from charged particles is expected.
For this mass region the track reconstruction method using the highest
efficiency, namely 99.9\% is used.  In this method a track is kept if
there are hits in two hodoscopes and one straw tube chamber.  The
efficiency for not rejecting a neutral particle is determined to be
about 61\%.  

For neutral strangelet candidates with baryonic mass greater than 30
contamination is a minor problem, therefore a track reconstruction
method is
used that emphasizes the elimination of ghost tracks that would increase
the rate of false elimination of neutral hits.  In this procedure hits
are required in all three hodoscope planes and one straw tube station.
In addition the time ordering of hits in the hodoscope had to be correct
and a $\chi^{2}$ cut on the track reconstruction is made if more than
one track shares a hodoscope hit.  The charge rejection efficiency is
determined to be approximately 97\%.       

In both of the track finding methods described above tracks are not 
required to originate from the target since they can result
from production of secondary particles.  Energy clusters with a matching
track are considered to be produced by charged particles and are 
discarded.  The masses of the remaining candidates are calculated using
the expression 

\begin{equation}
m = \frac{1.1E_{3x3}}{\gamma-1}
\label{eq:calmass}
\end{equation}

where $\gamma=(1-\beta^{2})^{-1}$.  $\beta$ is determined from a straight 
line path from the target to the peak tower and the time measured by the
peak tower.  The factor 1.1 accounts for partial shower containment in
the $3x3$ array of towers. 

\subsection{Contamination of neutral candidates}

Contamination of neutral cluster candidates by extra energy from
neighboring clusters or late hitting particles can imitate a high mass
object so that light particles such as protons or neutrons can be
misinterpreted as heavier particles or strangelets.  Time contamination
can result from particles produced in interactions closely spaced in
time in the target or particles produced in secondary interactions in or
upstream of the target and delayed relative to the triggered
interaction.  Many double beam events are rejected by eliminating those
that correspond to two Au ions traversing the quartz plate of the
MITCH counter during it's ADC integration time.  Some particles produced
in secondary interactions are identified and rejected using the
interaction veto counters located just upstream from the target.
Particles from secondary interactions are also eliminated by a cut on
the time the particles left the target.  Every event that had at least
one track generated later than 2.5 ns after the event start time is
rejected.  Events are also rejected that contained photons whose time 
intercept at the target exceeds 3 ns relative to the start time of the 
event.  Photons are identified by their narrow calorimeter showers where
typically 
the peak tower accounts for more than 95\% of the total shower energy. 

False reconstruction of heavy particles can also be caused by energy
contamination due to overlaps of two or more particle showers.  A shower
is considered to be contaminated if there are significant deviations
from the lateral energy profile and time distribution of a reference
shower.  The reference energy profile is constructed from a sample of
several thousand well isolated clusters matching tracks identified as
protons, deuterons or tritons.  Clusters are rejected if the energy
measured by the eight neighbor towers to a peak tower exceeds a maximum
fractional energy prescribed by the shower shape.  The maximum
fractional energy is chosen so as to achieve a 98\% efficiency per
tower.  Clusters are also rejected if the time measured by any of the
eight nearest neighbor towers differ significantly from the time
measured by the peak tower.  Further details on the analysis are given
in \cite{munhoz}.

\section{Production Limits}

\subsection{Calculation of limits}

Production limits can be calculated from the expression given below 
for the number of candidates observed ($N_{obs}$) as a function of 
spectrometer acceptances and efficiencies ($\epsilon$) in various regions 
of rapidity (y) and transverse momentum ($p_{\perp}$).
In the expression for $N_{obs}$, $\sigma_{c}$ is the strangelet
production cross section for 10\% central interactions (10\% of the
total cross section), $I_{c}$ is the number of central interactions
examined, $\epsilon(y,p_{\perp})$ is the efficiency for detecting a
strangelet as a function of y and $p_{\perp}$ and
$d^{2}\sigma/dydp_{\perp}$ is
the strangelet differential cross section.

\begin{equation}
N_{obs} = \frac{I_{c}}{\sigma_{c}}\int\epsilon(y,p_{\perp})
\frac{d^{2}\sigma}{dydp_{\perp}}dydp_{\perp}
\label{eq:candid}
\end{equation}

In order to set total production limits for strangelet production it is
necessary to have a model for the production of strangelets as a
function of phase space.  Then this model is integrated over the limits
in each phase space bin to obtain the final limit.  We assume a 
strangelet production model separable in y and p$_{\perp}$:

\begin{eqnarray}
\frac{d^2\sigma}{dp_{\perp}\; dy}  & \propto & 
\left [ p_{\perp} \; \exp{\left ( \frac{-2p_{\perp}}{<p_{\perp}>} \right
)} \right ] \;
\left [ \exp{\left ( \frac{-(y-y_{cm})^2}{2\sigma_{y}^2} \right ) } 
\right ] 
\label{eq:model}
\end{eqnarray}

where $\sigma_{y}$ is the RMS width of the rapidity distribution and 
$<p_{\perp}>$ is the mean transverse momentum of the strangelet.  In
order to calculate the total acceptance and efficiency we use a rapidity
width $\sigma_{y}$ of 0.5.  The rapidity and transverse momentum
distributions were assumed to be uncorrelated.  This production model has
been widely used in strangelet searches \cite{strange,ambro}.  

\subsection{Determination of limits for charged strangelets} 

The first task in the strangelet search is to use the time of flight 
and reconstructed momenta associated with the tracks with appropriate 
cuts to establish a set of high mass candidates.  At this stage of the
analysis a large number of high mass candidates are always seen.  This
is due to charge exchange scattering of neutrons discussed above that 
produces tracked protons with reconstructed momenta that are too large.
Most of these give calorimeter masses near that of the proton.

Using the efficiencies determined for observing strangelets, the upper
limits on their production can be determined.  The final limits are 
quoted as 90\% confidence level limits in 10\% most central interactions
of 11.5 GeV/c per nucleon $Au$ projectiles with $Pb$ or $Pt$ targets.  
The limit is given as:

\begin{equation}
90\% C.L. = \frac{N_{Poisson}}{N_{sampled}\epsilon_{accept}\epsilon_
{tracking}\epsilon_{calorimeter}\epsilon_{trigger}}
\label{eq:cl}
\end{equation}

The 90\% confidence level limit from Poisson statistics is $N_{Poisson}
= 2.30$ and $N_{sampled}$ is the total number of events sampled.  

The efficiencies in the 90\%C.L. formula above vary both with
strangelet species (A,S) and with the production model.  Below 
representative values are given.  The overall geometric acceptance 
$\epsilon_{accept}$ is approximately 8\%.  The tracking efficiency 
$\epsilon_{track}$ including track quality cuts is approximately 75\%.
The calorimeter contamination cut efficiency $\epsilon_{calorimeter}$ 
varies over a large range of from 40\% to 80\% depending on the incident
particle occupancy.  The trigger efficiency $\epsilon_{trigger}$ is
high varying from 90\% to 100\%.  

The above efficiencies are calculated using a full GEANT
simulation of the experiment that includes magnets, vacuum chamber,
detectors, etc.  Detector survey data is used as input for the detector
geometries and GEANT calculates the geometric acceptance and single
particle tracking efficiency.  The efficiency of a given detector is
determined by using the data to find tracks in the other detectors and
then checking for a consistent hit in the detector.  In order to
determine multi-track efficiencies and calorimeter shower cut
efficiencies, Monte Carlo detector hit information which simulates the
measured detector responses is overlayed with real experimental data.
The results are then processed through our tracking and shower
analysis.     

\subsubsection{Limits for positively charged strangelets}

\begin{figure}
\psfig{file=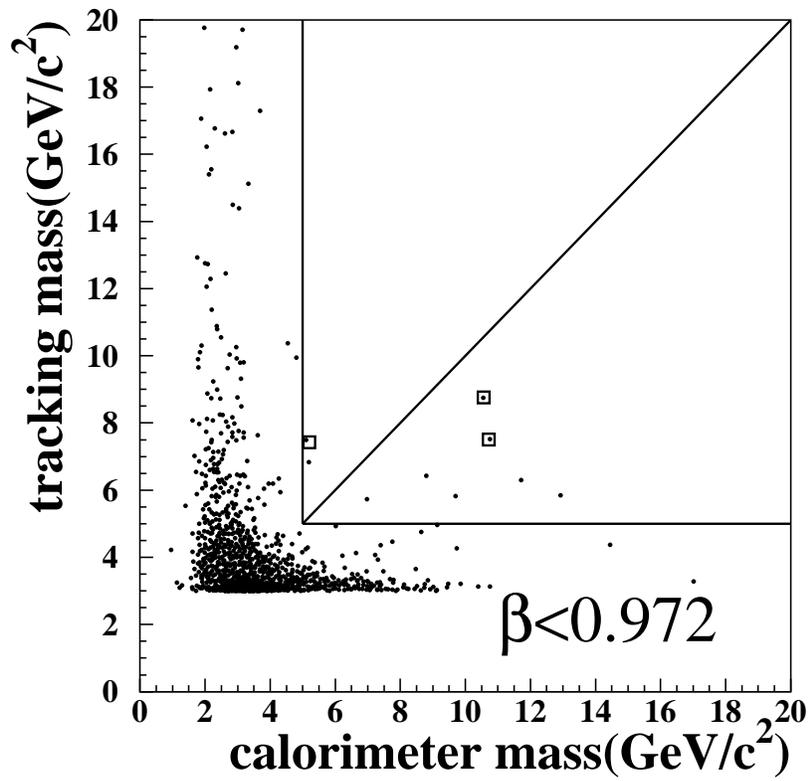,width=5.0in}
\caption{Tracking mass vs. calorimeter mass distribution of charge=+1
candidates with $\beta\leq{0.972}$.  Data points in the rectangles are
for those candidates with mass greater than $5 GeV/c^{2}$.}
\label{fig:Z=1}
\end{figure}

In order to determine limits on the production of positively charged
strangelets a total of 13 billion of the 10\% most central events are
sampled.  In order to search for strangelets the masses of candidates
identified in the tracking process are matched with the corresponding
masses measured in the calorimeter.  A number of possible candidates
are observed with a loose cut of $\beta\leq{0.985}$.  A tighter cut of
$\beta\leq{0.972}$ results in a cleaner spectrum as well as better mass
resolution.  The corresponding plot of calorimeter vs tracking mass for
Z=+1 high mass candidates is shown in Figure~\ref{fig:Z=1}.

As can be seen in Figure~\ref{fig:Z=1} there are a handful of candidates
with rough agreement between calorimeter and tracking mass.  Next a cut
is made on the consistency of the kinetic energy as measured in the
calorimeter and by tracking.  Only three candidates with both tracking 
and calorimeter masses greater than $5 GeV/c^{2}$ survive all cuts
including the kinetic energy cut.  These three candidates indicated by
squares in the figure were examined in great detail.  In each of these
there are several towers with energy deposited greater than 1 GeV but 
with no timing information.  For hits later than a preset time no timing
signal is given by the calorimeter.  This implies that these events
are contaminated by a second interaction from a late hit in the target.  
Events are rejected if they contain towers with energy greater than 1
GeV but no timing information.  On this basis the three candidates are 
thus judged to be consistent with background.  The efficiency of the
above cut is 85\%.  A detailed discussion of this analysis is given by
Xu\cite{xu}.

A search was made for heavy objects with Z=+2.  From
Figure~\ref{fig:zxb2} with the tight $\beta$ cut it is clear that we see
a peak due to $^{6}He$ but no candidates above mass 6.  $He$ isotopes
with mass 5 and 7 are unstable
against prompt particle emission but $^{8}He$ with a halflife of 119 ms
would be observable.  From Figure ~\ref{fig:zxb2} it is evident that no
mass 8 events are observed.  

\begin{figure}
\centerline{\psfig{file=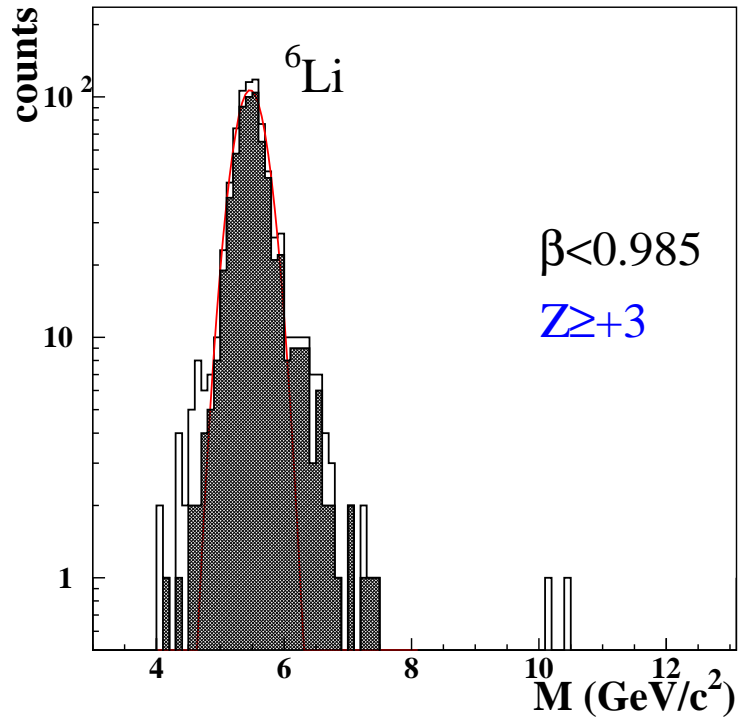,height=6.0in}}
\caption{Mass distribution for Z=+3 strangelet candidates}
\label{fig:Z=3}
\end{figure}

It is possible to identify particles with $Z\geq{3}$ but distinguishing 
between Z = 3 and higher is difficult due to saturation of the hodoscope
ADCs.  The corresponding plot for high mass candidates with $Z\geq{3}$
is shown in Figure~\ref{fig:Z=3}.  The cuts are the same as those
applied to Z = 1 and 2.  The peak is identified as $^{6}Li$ with a high 
mass shoulder from $^{7}Li$.  Note that the two candidates in the figure 
between mass numbers 10 and 11 were eliminated by the tight calorimeter 
cut.  The conclusion is that there are no
strangelet candidates with $Z\geq3$ and $m\geq8 GeV/c^{2}$.  

\begin{figure}
\centerline{\psfig{file=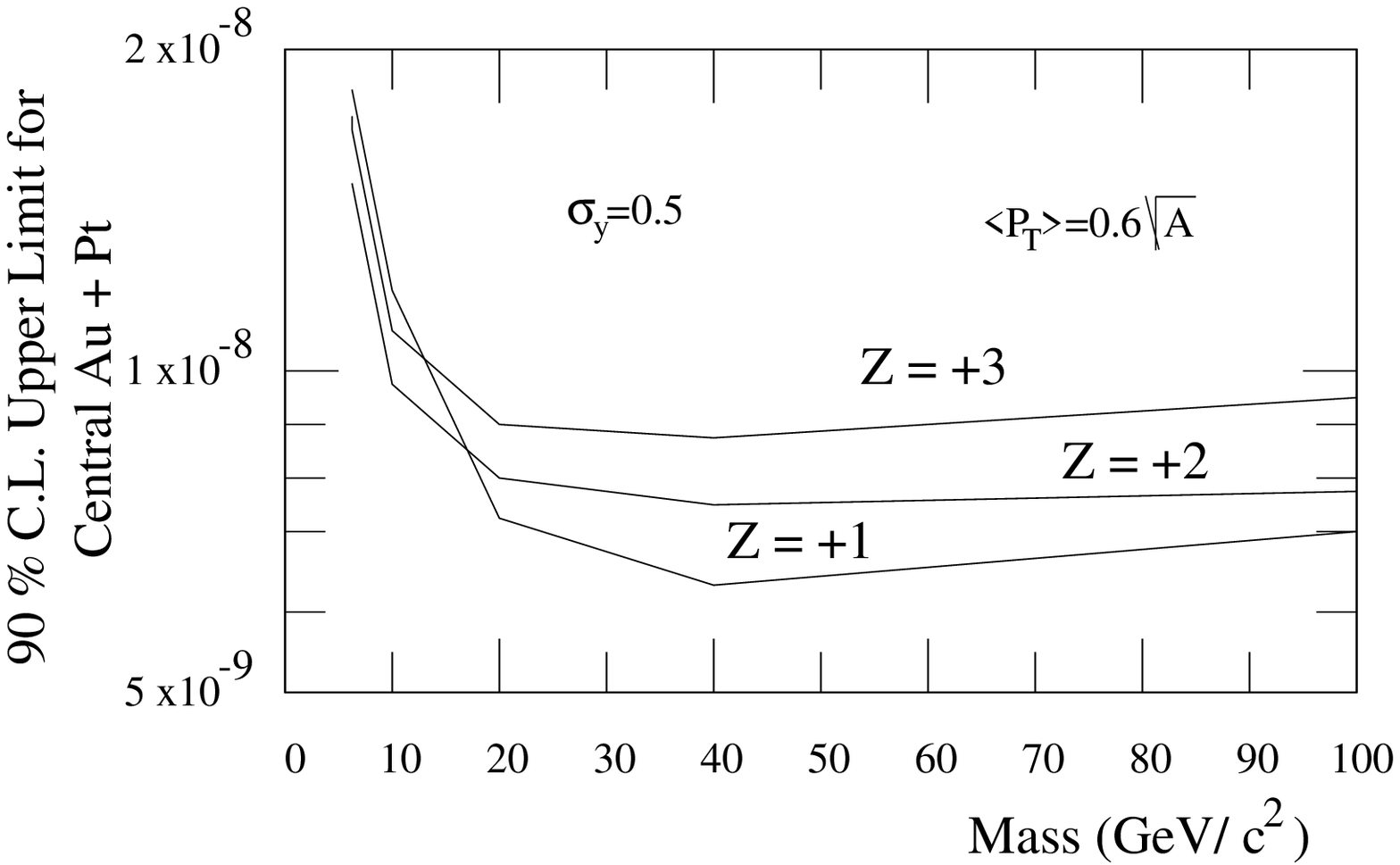,width=6.0in}}
\vspace{+0.5in}
\caption{90\% C.L. upper limits for the production of strangelets with
positive charges per central $Au$ + $Pt$ collision at a beam momentum of
11.5 GeV/c per nucleon.}
\label{fig:poslimits}
\end{figure}

Based on the null results of the searches for positively charged
strangelets with Z = 1, 2 or 3 we can set limits at $90\%$ C.L.
over a wide mass range for production of strangelets from the
interaction of 11.5 GeV/c per nucleon $Au$ projectiles with $Pt$
targets.  These limits are shown in Figure~\ref{fig:poslimits} and the 
corresponding numerical values are shown in Table~\ref{tab:pos_limits}.
A total of 13 billion $10\%$ most
central interactions are sampled.  The limits are below $2\times10^{-8}$
per central interaction and are relatively constant above a
mass of 8 GeV/$c^{2}$. 

\subsubsection{Limits for negatively charged strangelets}

\begin{figure}
\centerline{\psfig{file=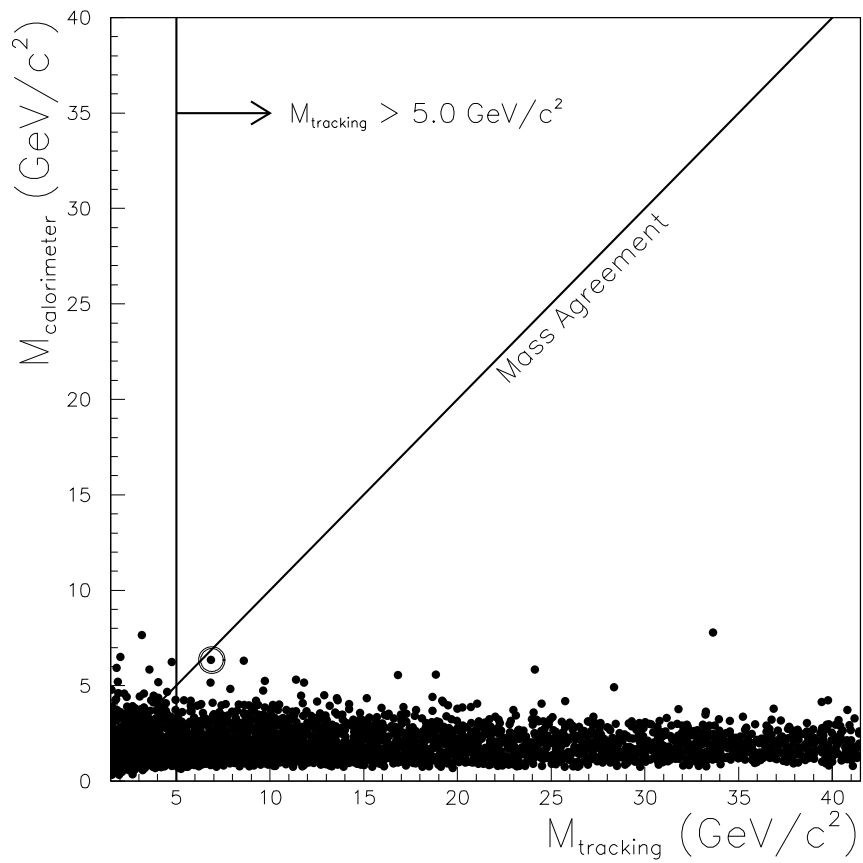,height=5.0in}}
\vspace{+0.5in}
\caption{Charge -1 strangelet candidate distribution in calorimeter vs
tracking mass.  the cut at 5.0 GeV/$c^{2}$ for the minimum tracking mass
is shown.  The single candidate whose tracking and calorimeter mass
agrees is circled}
\label{fig:Z=-1}
\end{figure}

In order to determine limits on the production of negatively charged
strangelets a total of 13.8 billion of the 10\% most central events were
sampled.  A number of cuts are applied to the tracking data as well as
the calorimeter data \cite{vanburen}.  To be considered a strangelet
candidate the mass from tracking is restricted to greater than 
5 GeV/$c^{2}$.  Application of these cuts results in a sample of 26,959
candidate tracks.  In order to search for strangelets, the masses of 
candidates identified in the tracking process are matched with the 
corresponding masses measured in the calorimeter.  The resulting 
distribution of tracks is shown in Figure~\ref{fig:Z=-1}.  In the figure
a large number of tracks are seen corresponding to large tracking masses
but small calorimeter masses.  As described above, these tracks are
believed to be mostly due to neutrons that charge exchange scatter and
thus masquerade as high mass particles.   

It is apparent from Figure~\ref{fig:Z=-1} that there are no good 
candidates
with masses above 10 GeV/$c^{2}$.  Below 10 GeV/$c^{2}$ the requirement
is made that the calorimeter energy match the tracking kinetic energy
within $-1\sigma$ and $+3\sigma$, where $\sigma$ is the energy
resolution of the calorimeter.  This final agreement cut is $84\%$ and
leaves only one candidate which is circled in Figure~\ref{fig:Z=-1}.
Some background processes have been identified that could fake such a
particle as discussed by Van Buren\cite{vanburen}.  Thus with only one
candidate it is not possible for us to determine if it is a strangelet
or background.

\begin{figure}
\centerline{\psfig{file=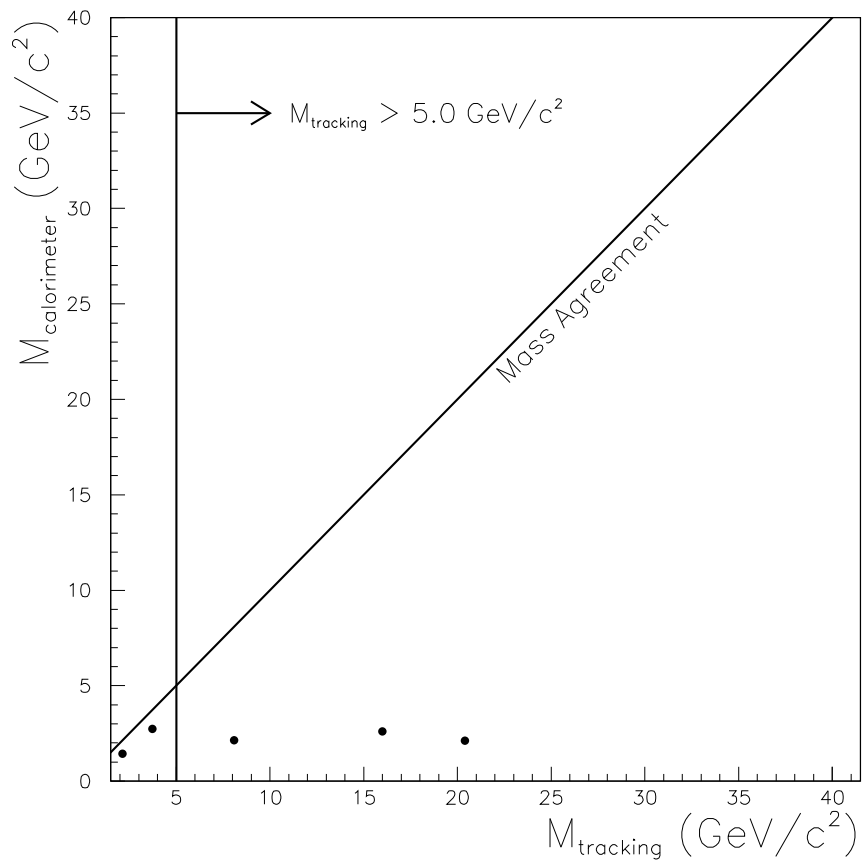,height=5.0in}}
\caption{Charge -2 strangelet candidate distribution in tracking vs
calorimeter mass}
\label{fig:Z=-2}
\end{figure}

A search was also made for strangelet candidates with charge of -2.  In
this case the same tracking cuts used for the Z=-1 case are employed.
No calorimeter cuts are used.  Above 5 GeV/$c^{2}$ in mass only 3
candidates are seen and none has a calorimeter mass near to the
tracking mass.  The results are shown in Figure~\ref{fig:Z=-2}.  The
efficiencies used in the Z=-2 analysis are discussed in \cite{vanburen}.

Based on the null results of the searches for negatively charged
strangelets with Z = -1 and -2 limits at $90\%$ C.L. are set over a 
wide mass range for production of strangelets from the interaction of 
11.5 GeV/c per nucleon  $Au$ projectiles with $Pt$ targets.  
Representative numerical
values for these limits are given in Table~\ref{tab:neg_limits}.  A
total of 13.8 billion $10\%$ most central interactions are sampled
using a negative B field from the analyzing magnet.  
If we assume that the candidate at A=7 for Z = -1 is a strangelet then 
the $90\%$ C.L. is increased by a factor of about 1.7.  The limits for 
the production of Z=-2 strangelets are based on a null result.  

\begin{figure}
\centerline{\epsfig{file=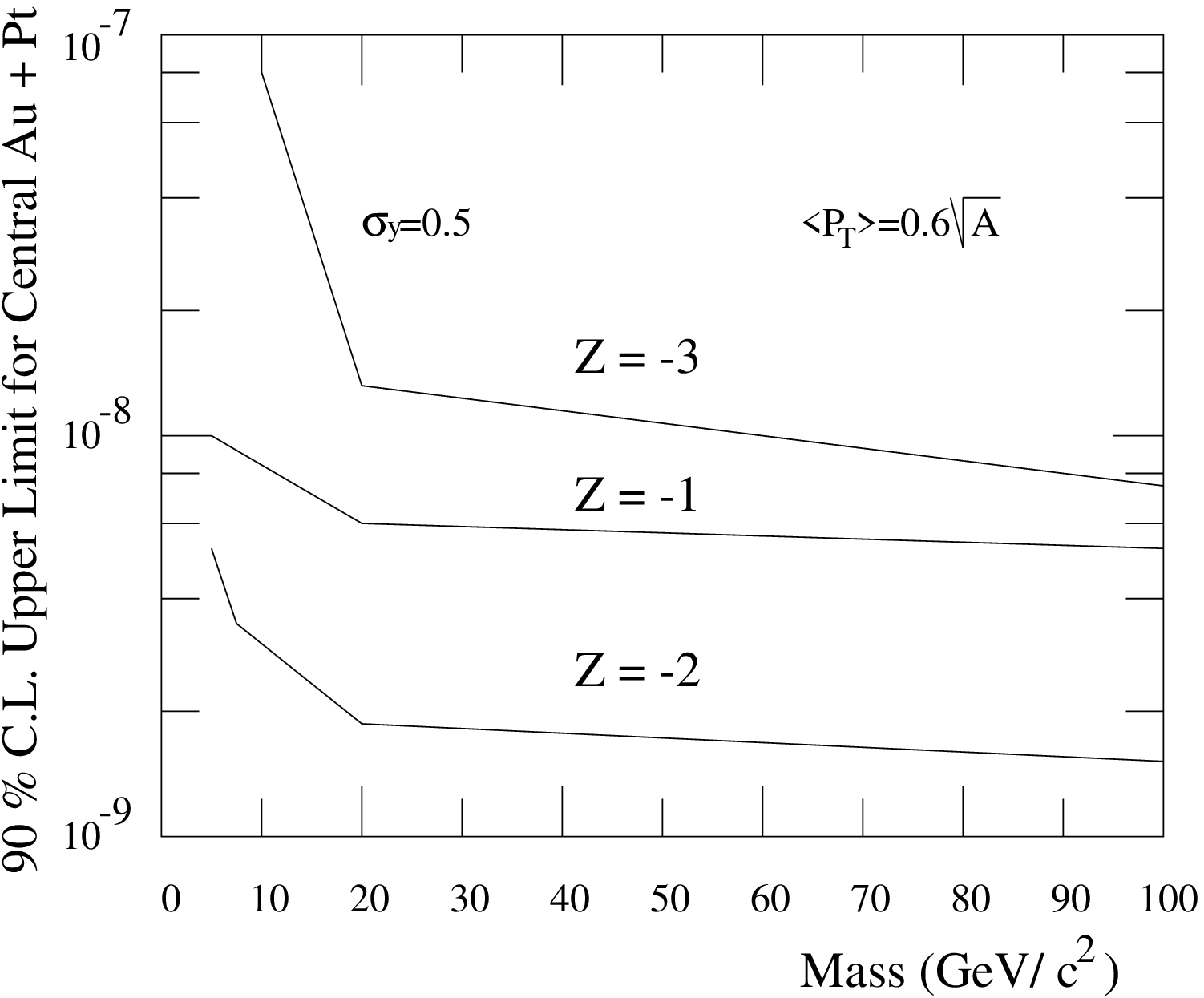,height=5.0in}}
\vspace{+0.5in}
\caption{90\% C.L. upper limits for the production of strangelets with
negative charges per central $Au$ + $Pt$ collision at a beam momentum of
11.5 GeV/c per nucleon.}
\label{fig:neglim}
\end{figure}

In addition to the above analysis the 13 billion 10\% most central
interactions observed using a positive B field from the analyzing magnet
and sampled in the search for positively charged strangelets were also
searched for negative strangelets.  In the analysis of this data set it
is also possible to search for strangelets with Z=-3 as well as Z=-1
and -2.  The limits determined by combining the results from the two
data sets are given in the last column of Table~\ref{tab:neg_limits} and
shown in Figure~\ref{fig:neglim}.

\subsection{Limits for neutral strangelets}

\begin{figure}
\epsfig{file=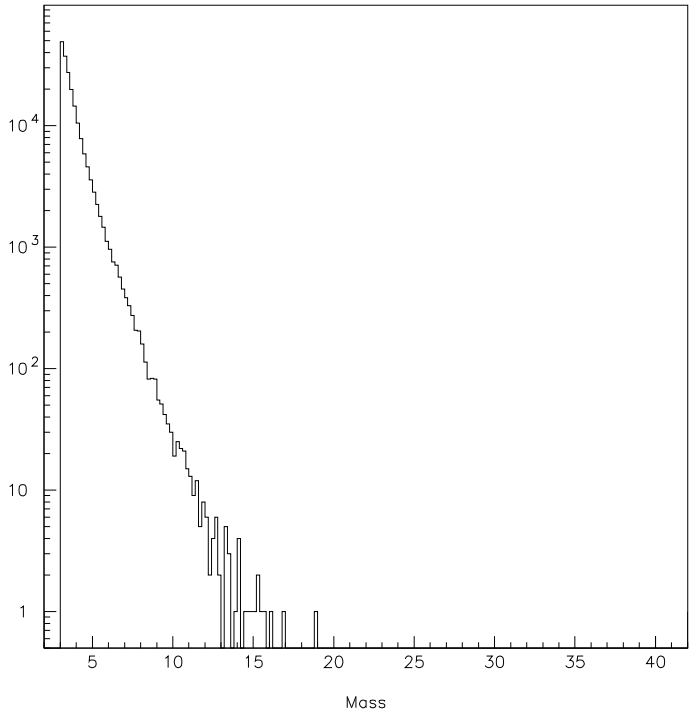,height=6.0in}
\vspace{+0.5in}
\caption{Reconstructed mass spectrum of neutral particle candidates.}
\label{fig:neutral}
\end{figure}

In order to determine limits on the production of neutral strangelets a 
total of 14.75 billion of the 10\% most central events are sampled.  A
reconstructed mass spectrum is shown in Figure~\ref{fig:neutral}.  For
the high mass region of the spectrum above 20 GeV/$c^{2}$ no candidates
are observed.  In the mass range from 3 GeV/$c^{2}$ to 20 GeV/$c^{2}$
there are 195721 candidates distributed roughly exponentially with
respect
to mass.  A detailed analysis \cite{munhoz} of the event structure of
the candidates and also candidates rejected by various contamination
cuts show that delayed upstream interactions are mainly responsible
for the higher mass candidates while lower mass candidates are mostly 
due to energy contamination from overlapping showers.  

\begin{figure}
\psfig{file=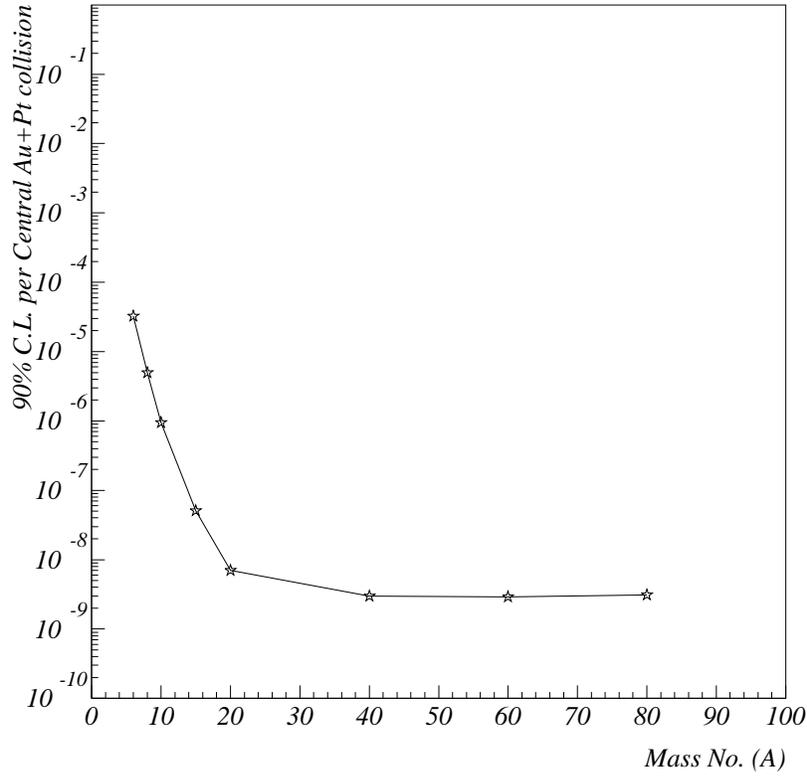,width=5.0in}
\vspace{-0.5in}
\caption{90\% C.L. upper limits for the production of neutral
strangelets per central $Au$ + $Pt$ collision at a beam momentum of 
11.5 GeV/c per nucleon.}
\label{fig:neutlim}
\end{figure}

The calculation of production limits for neutral strangelets therefore
proceeds based on the number of candidates observed.  For masses above
20 GeV/$c^{2}$ no candidates are observed.  For masses below 20
GeV/$c^{2}$ the sensitivity is limited by overlapping showers and
double interactions not vetoed by the electronics.  We assume no 
knowledge of
the background and no restriction in the production of strangelets.  The
number of observed candidates as shown in Figure~\ref{fig:neutral} in
the mass range $m\pm1.25\sigma_{m}$ is used to estimate the 90\% C.L. 
upper limit for production of neutral strangelets which is shown in
Figure~\ref{fig:neutlim} and Table~\ref{tab:neut_limits}.  As can be
seen from Figure~\ref{fig:neutlim}
the limit is nearly flat above 20 GeV/$c^{2}$ due to the large
acceptance of the E864 spectrometer.  The lower sensitivity at lower
masses is due to background contributions.    

\subsection{Limits on $H^0$ production}

The large neutral background at low mass in this experiment makes a
direct search for the $H^0$ impractical.  In addition such a search
would have only been sensitive to $H^0$s with proper lifetimes greater
than about 50 ns unlike previous searches which typically had no such
restriction.  A search was made for the $H^0-d$ hybrid bound state of
the $H^0$ and the deuteron.  The
$H^0-d$ would have Z=+1 and mass between an $\alpha$ particle and the 
mass of $d+\Lambda\Lambda$.  Assuming a $H^0-d$ mass of 4.09 GeV/$c^2$, 
the background in this mass region is dominated by
the triton tail.  Using a tighter rapidity cut of $y\leq1.9$ to clean up
the spectrum, no significant peak is observed around the $H^0-d$ mass.  

A detailed discussion of the analysis leading to the $H^0-d$ limit has
been given by Xu\cite{xu}.  A mass resolution of 2\% from the
triton mass peak and a double-exponential fit to the triton 
tail is used to determine the upper limit.  Given the fact that there 
is no particle with Z=+1 around a mass of 4.1 GeV/$c^2$ and the
excellent fit to the triton tail, a 90\% C.L. limit for $H^0-d$
production of $0.92\times10^{-7}$ per 10\% most central collision is
obtained.   

Baltz\cite{baltz} estimated that for central and min-bias $Au$ + $Au$
collisions at AGS energies the predicted number of bound
$\Lambda\Lambda$ particles is 0.012 and 0.07 per collision,
respectively.  Using the penalty factor of 48 for the addition of one 
nucleon by coalescence as measured in E864 (see discussion in the next 
section) and the limit for the $\Lambda\Lambda$ of 0.012 we obtain a 
predicted production level for the
$H^0-d$ of $5\times10^{-6}$.  This is a factor 54 times higher than our
measured limit.  The proper lifetime for a particle in the E864
spectrometer is about 50 ns so the above result indicates that it is
unlikely that the $H^0$ exists with a lifetime greater than about 10 ns.
Nevertheless the $H^0$ could exist with hyperonic lifetimes down to
about $10^{-10}$ seconds.

\subsection{Limits on production of pineuts}

Pineuts are hypothetical bound states of a negative pion with two or 
more neutrons.  A number of authors have speculated on the existence of
such states \cite{gale,kalb,eric,frei,garc}.  Pineuts might exist as a 
result of the attractive $\pi$-N interaction.  A $\pi$-2n bound state 
would for instance have a mass of around 2019 MeV could only decay via 
weak interactions, since there is no negatively charged nucleon.
Pineuts might therefore have lifetimes of the order of the lifetime of 
charged pions.  Such objects if produced in heavy ion reactions would be
readily observed in the magnetic spectrometer of the E864 experiment as 
heavy objects with Z=-1.  The $\pi$-2n could be found as a mass peak 
between the $\overline{d}$ mass and 2019 $GeV/c^{2}$.

\subsubsection{Past searches for pineuts}

Experimental searches for pineuts were first conducted using light ion 
collisions with negative results \cite{boer,ashe,will,park}.  Searches 
were also performed using heavy ion collisions of $^{40}Ar$ and 
$^{139}La$ projectiles at the Bevalac at kinetic energies of 
1.8 GeV/nucleon and 1.26 GeV/nucleon, respectively, incident on targets 
of $^{238}U$ \cite{park,fwn}.  Projectiles of 14.6 GeV/nucleon $^{28}S$ 
from the AGS, on $Pb$, $Sn$, $Cu$ and $Al$ targets \cite{barr} and 
100 MeV/nucleon $^{18}O$ 
projectiles at RIKEN \cite{suzu} on $Be$ targets were used in pineut 
searches.  Heavy ion collisions at high energies provide a unique 
environment for the production of pineuts given that in these collisions
large quantities of pions are produced and can, in principle, combine to
the numerous neutrons already present in the projectile and target 
nuclei.  A recent calculation using a coalescence model with the event 
generator ARC \cite{kahana} predicted that pineuts, should they exist, 
would be produced at detectable levels in high-energy heavy ion 
interactions.  However, the search conducted at the AGS by the E814 
collaboration using $^{28}Si$ projectiles obtained an upper limit on 
pineut production of $10^{-6}$ per collision in contrast to the 
prediction of the ARC based dynamical coalescence calculation of a
production level of $10^{-3}$ per collision.  It is relevant to note
that coalescence calculations based on ARC typically underestimate the 
penalty factors for the production of composite objects such as 
deuterons, and other light nuclei\cite{light}.  Also the probability for 
producing weakly bound pineuts might be further reduced by final state 
interactions in heavy ion collisions.  

\subsubsection{Measurement, analysis, results and conclusions}

Pineuts produced at central rapidities, in central $Au$ + $Pt$
collisions with lifetimes in access of that of the charged pion have a 
finite probability of reaching the calorimeter located at the end of the
E864 spectrometer.  They would produce high rigidity tracks that could
be reconstructed using the same techniques used for strangelets.  We
therefore extend our negative strangelet search to look for mass peaks
at pineut masses of 2019 $MeV/c^{2}$, 2957 $MeV/c^{2}$, etc.

The low mass region suffers from reduced trigger efficiencies as well as
significant background from scattered protons and $\overline{p}$.  In 
contrast, antinuclei of similar masses deposit annihilation energy in
the calorimeter therefore improving the trigger efficiency as well as 
making it easier to distinguish background.  No signal was observed for
the A=2 pineut ($\pi-2n$) thus an upper limit on it's invariant yield of
$2.5\times10^{-7}$ per 10\% most central ccollision near mid rapidity is
significantly higher than the observed $\overline{d}$ signal of 
$3.7\times10^{-8}$ \cite{dbar}.  The corresponding upper limits for the 
$\pi-3n$, $\pi-4n$ and $\pi-5n$ states are found to be
$7.0\times10^{-8}$, $2.5\times10^{-8}$ and $1.5\times10^{-8}$, 
respectively.

We have searched for the production of pineuts in collisions of $Au$
beams with a target of $Pt$.  We find no evidence for the production of 
pineuts at a sensitivity level which surpasses both that of previous 
studies and the predictions of a dynamical coalescence model.  This 
analysis confirms earlier studies \cite{barr} that such particles are
not likely to have pionic lifetimes.

\section{Comparison with and Constraints on Strangelet Production
Models}

A goal of this experiment is to either discover SQM or to use the 
measured limits to make some statement concerning the stability of SQM 
and constrain the bag model parameters that predict metastable strangelets
in the mass and lifetime range studied.  An additional complication is
the fact that mechanisms by which strangelets can be produced in
relativistic heavy ion collisions are not well known so different
production models need to be considered.  Below we examine both plasma
and coalescence production models in light of the production limits
measured in this experiment.

\subsection{Constraints on plasma production models}

Greiner \emph{et al.} \cite{greiner} suggest a mechanism for
strangelet formation involving the formation of a Quark-Gluon Plasma
(QGP) followed by the emission of a strangelet.  In this scenario
$\overline{s}$ quarks produced in a baryon rich QGP combine with
abundant $u$ and $d$ quarks to form K mesons.  This ``strangeness 
distillation'' could result in a residue rich in $s$ quarks from which 
strangelets might form during the cooling process.  In this scenario it 
might be possible to produce large strangelets with $A \geq 15$.  
Nucleation calculations carried out by Kapusta \emph{et al.} 
\cite{kapusta} predict that under certain conditions a QGP might be 
formed in as many as 1 in 100 to 1000 central collisions.  

Using the production limits obtained for charged and neutral strangelets
it is not possible to rule out any of the individual steps in the above 
QGP distillation scenario but it is possible to place limits on the
overall process.  We define BF(QGP) to be the branching fraction for the
formation of the QGP from interaction of 11.5 GeV/c per nucleon $Au$ 
projectiles with a $Pt$ or $Pb$ target and BF(Strange) to be the 
branching ratio for decay of the QGP into a strangelet.  Then a model 
independent upper limit can be set on the product of these two
processes.
  
\begin{figure}
\epsfig{file=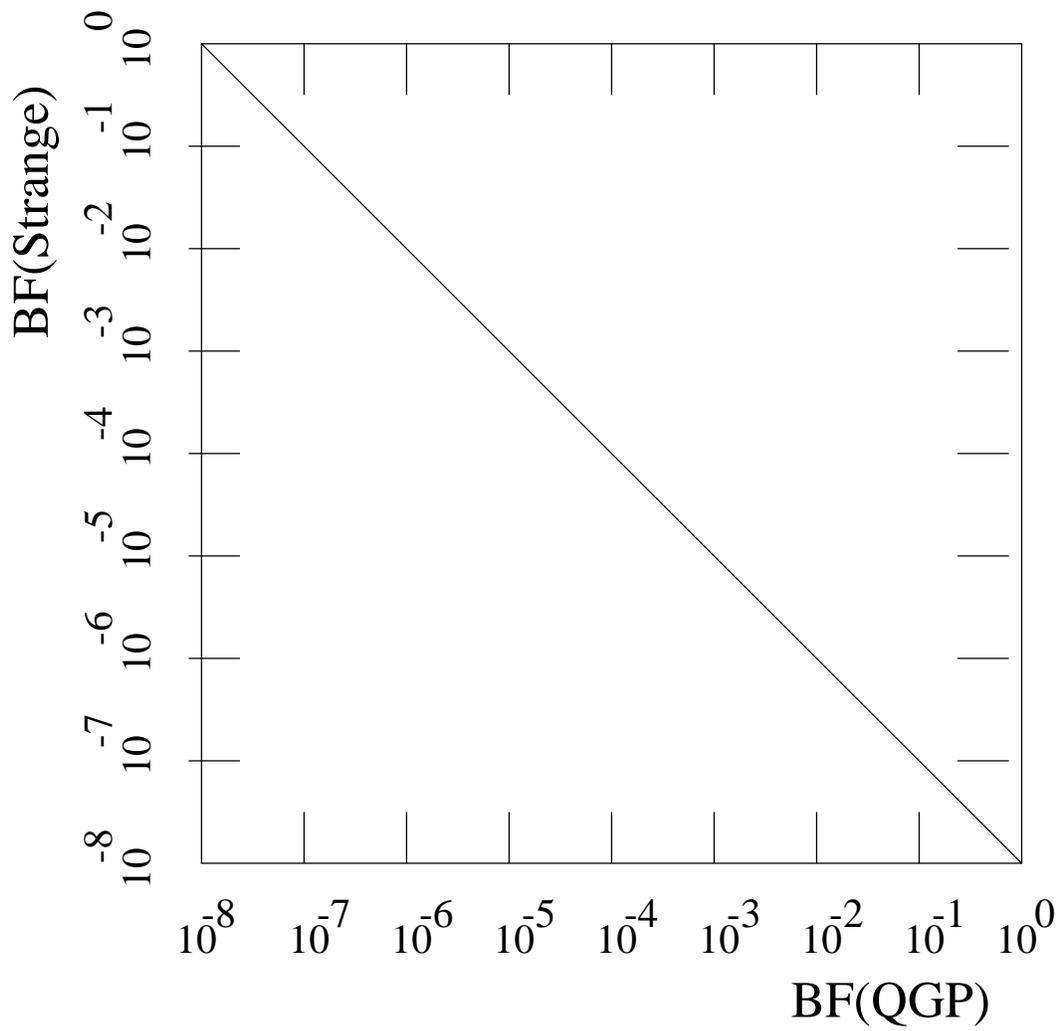,width=5.5in}
\caption{Typical branching fraction limits for distillation of charged
strangelets from a QGP}
\label{fig:qgpcharge}
\end{figure}

For charged strangelets the production limits are relatively flat as a
function of mass.  As an example we consider a typical production limit
for a charged strangelet of given A and Z of $1 \times 10^{-8}$ for 10\%
most central collisions at the 90\% C.L.  The corresponding limits as a 
function of BF(QGP) and BF(Str) are
shown in Figure~\ref{fig:qgpcharge}.  The numbers can be refined for
charged strangelets of a given mass and charge using the information
from Table~\ref{tab:pos_limits} or Table~\ref{tab:neg_limits}.  As an 
example if the QGP was produced in 1 collision per 1000 central 
collisions, the probability that a Z=+2 strangelet with A=10 would be 
produced upon cooling of the QGP is less than 0.001\%.

Crawford \emph{et al.} \cite{crawford} have made specific predictions 
concerning production rates of strangelets following a QGP phase
transition.  Their model assumes formation of a QGP in every 10\% most
central collision.  In the model the large QGP drop fragments into
smaller QGP droplets and then can cool primarily by meson emission to 
form an S drop with a given A and Z.  Finally this S drop can cool
partially by gamma emission to form a strangelet of given S, A and Z.
The probabilities of the above sequence of events leading to formation
of a strangelet with lifetimes greater than $3 \times 10^{-8}$ sec is
calculated for $\gamma_{lab}$ of 14.5, 60 and 200.  The predictions are
for strangelet mass numbers (A) of 10, 15 and 20 with charges (Z)
ranging from -4 to +4 and are given in Table VI of 
Reference~\cite{crawford}.  For 
$\gamma_{lab}$ of 14.5 that is most relevant for this experiment the
two highest probabilities are $4.8 \times 10^{-7}$ for A=10 and Z=3 and
$7.5 \times 10^{-8}$ for A=10 and Z=2.  The production limits of $1.1
\times 10^{-8}$ and $9.8 \times 10^{-9}$ shown in
Table~\ref{tab:pos_limits} from this experiment therefore test the 
Crawford \emph{et al.} limits under their assumption of the production 
of a QGP droplet in every central collision.

\begin{figure}
\epsfig{file=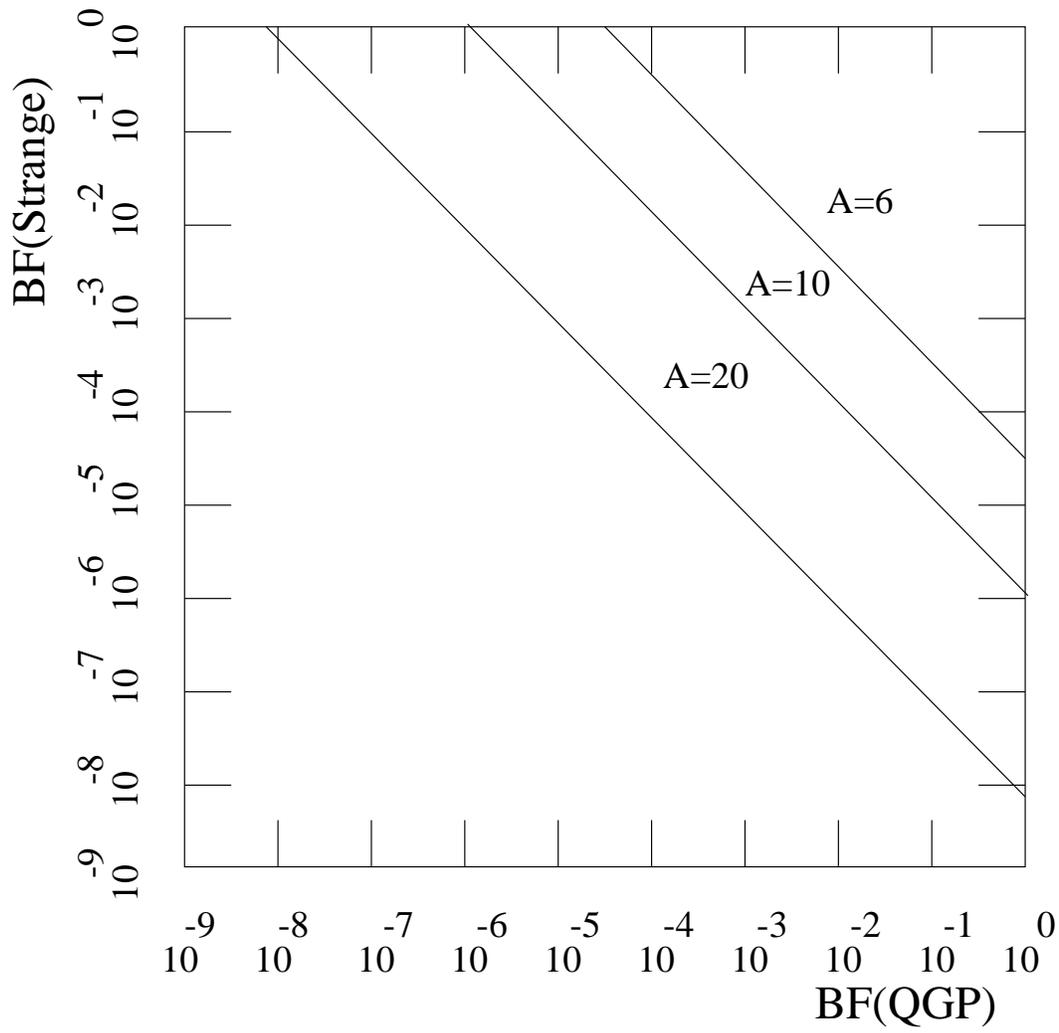,width=5.5in}
\caption{Branching fraction limits for distillation of neutral
strangelets from a QGP.}
\label{fig:qgpneut}
\end{figure}

It is also possible to estimate limits for the production of neutral
strangelets under the above scenario.  The results are given for
strangelets with A = 6(quark-alpha), 10 and 20.  The corresponding 
branching fractions are shown in Figure~\ref{fig:qgpneut}.  The
production limits determined in the Crawford model for neutral 
strangelets are 
$2.5 \times 10^{-8}$ and $2.6 \times 10^{-9}$ for A equal 10 and 15,
respectively, so our sensitivity is not great enough to test this model
for neutral strangelets.     

\subsection{Constraints on coalescence production models}

A very different production mechanism for strangelets involves the
coalescence of strange and non-strange baryons produced in heavy ion
collisions.  In this picture, just after the collision the produced
particles undergo many interactions but after the system has expanded
significantly baryons that are close to each other in configuration and
momentum space may fuse together to form nuclei and
hypernuclei.  Hypernuclei have lifetimes of the order of the $\Lambda$
particle and do not traverse our spectrometer, but if a strangelet
state of similar quantum numbers (A,S) is more stable than the
hypernucleus, the hypernucleus could act as a doorway to the strangelet 
state.  

Baltz \emph{et al.} calculate the production rate of strange clusters in
relativistic heavy ion collisions using a simplified coalescence model 
and the ARC cascade code \cite{baltz}.  Of particular interest for this 
work are their predictions for hyperfragment production for central 
$Au$ + $Au$ collisions at AGS energies.  If it is assumed that a
strangelet of given A and S is produced at approximately the same rate
as a hyperfragment with the same A and S we can compare the calculated 
hyperfragment limits with our strangelet limits.  The most relevant
comparison is for strangelets with A = 6 and 7.  The calculated yield
for the $^{6}_{\Lambda\Lambda}He$ of $1.6\times10^{-5}$ is higher than
the experimentally measured limit for a Z = 1 and A = 6 strangelet of 
$1.8\times10^{-8}$.  Another relevant comparison is with the calculated
yield of the $^{7}_{\Xi\Lambda\Lambda}He$ of $2\times10^{-7}$ with our
measured limit for a Z = 2 and A = 7 strangelet of approximately 
$1.4\times10^{-8}$.  Thermal models predict production yields that are
below our sensitivity for low-mass strangelets.  For example, the rate
for $^{7}_{\Xi\Lambda\Lambda}He$ production is calculated to be 
$\sim2\times10^{-10}$ in $Au$ + $Au$ collisions\cite{pbm}.

It is also important to note that in our experiment $^{6}He$ but not
$^{8}He$ is observed ($^{7}He$ is particle unstable).  Since there are
additional penality factors associated with the addition of a unit of
strangeness it can be concluded that this experiment does not have the 
sensitivity to observe strangelets produced by coalescence with $A\geq8$
and is marginal for A=7.  On the other hand, observation of a charged
strangelet with $A\geq10$ could be a relatively clean signature for the 
formation of the QGP.

\section{Constraints on Future Searches}

\begin{figure}
\psfig{file=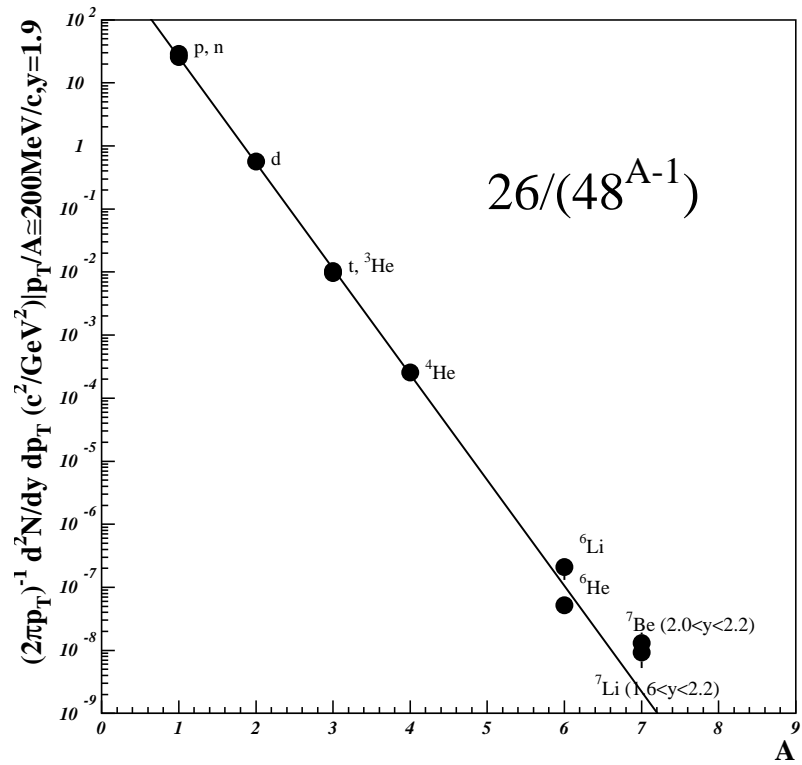,width=5.0in}
\vspace{-0.5in}
\caption{Invariant yields at or near y = 1.9 and $p_{\perp}/A$ = 
200 MeV/c as a function of mass number A.}  
\label{fig:finch}
\end{figure}

In addition to searching for strangelets, experiment E864 has carried
out a comprehensive set of measurements which address the coalescence of
multibaryon states in heavy ion collisions at AGS energies.  Production
of stable light nuclei by coalescence is observed from A=1 to A=7.  The
results for the stable light nuclei have been published\cite{light}.
The invariant yields for stable nuclei from A=1 to A=7 are shown in 
Figure \ref{fig:finch} for y near 1.9 and $p_{T}$/A = 200MeV/c as a
function of mass number.  As can be seen from the figure the addition by
coalescence of each nucleon involves a penalty factor of about 48.  As
an example, taking the
production of $^{6}He$ as $2\times10^{-7}$ per 10\% most central
collision, the probability for producing a strangelet with A=7, Z=2 and
S=-1 is $\leq 4\times10^{-9}$, which is below our limit for such
strangelets.  These results indicate that if such strangelets are formed
by coalescence then a search with a sensitivity of at least a factor of 
10 greater than in this experiment will be needed.  If even larger
strangelets are formed by coalescence, we'll need even a greater
increase in the sensitivity.

\section{Conclusions}

The E864 spectrometer is used to sample approximately 27 billion 10\%
most central $Au$ + $Pt$ interactions at the AGS in a search for charged 
strange quark matter.  In addition 14.75 billion 10\% most central 
$Au$ + $Pt$ interactions are sampled in a search for neutral strangelets.
Redundant tracking methods and calorimetry are used to reduce
background.  No consistent candidates for new states of strange quark
matter are found with proper lifetime greater than approximately
50 ns.  The search results in the assignment of 90\% C.L. upper limits
of typically $10^{-8}$ or less for 10\% most central
collisions of $Au$ + $Pt$ for charged strangelet searches over a mass
range from A = 6 to 100.  We also report here limits on the 
production of neutral strangelets.  The 90\% C.L. upper limit is 
$\leq10^{-8}$ for $A\geq20$ and increases to $10^{-6}$ for A=10.
Coalescence studies of light nuclei indicate a coalescence penalty
factor of about 48 for the addition of each nucleon.  An additional 
penalty factor may exist for replacement of a non-strange by a strange 
quark.  This is being investigated by studying the yield of the
$^{3}_{\Lambda}H$ in the E864 experiment and the result will be reported
in a forthcoming publication.

Although we are able to set very low upper limits on the existence of
strangelets in the range of sensitivity of our experiment we are not
able to answer the question concerning their existence.  There are
several definite reasons for this.  First the experiment is only 
sensitive to strangelets with proper lifetimes greater than about 50 ns.
Also the penalty factor for addition of a nucleon to a fragment
is found to be about 48 as shown in our results on the production of 
light nuclei from coalescence\cite{light} which is much higher than
expected.  The experiment is not sensitive enough to detect $^{8}He$, 
therefore detection of coalescence-produced strangelets with $A\geq8$ 
would not be expected.  It did rule out the formation of a QGP followed 
by the formation of a
strangelet at levels of typically $10^{-8}$ per 10\% most central
collision.  These studies represent the most extensive and highest
sensitivity heavy ion based searches at AGS energies for SQM to date.  
In addition high efficiency searches at the CERN-SPS by
NA52 \cite{borer,appel} found no evidence for SQM.  We conclude from 
these studies that if strangelets exist and can be produced in 
relativistic heavy ion collisions, experiments with very much higher 
statistics will be needed in order to detect them.  Nevertheless, if 
the QGP can be made this might produce additional pathways for 
strangelet production.   

\section{Acknowledgements}

We acknowledge the excellent support of the AGS staff.  This work was
supported by grants from the U.S. Department of Energy's High Energy and
Nuclear Physics Divisions, the U.S. National Science Foundation, the
Istituto Nationale di Fisica Nucleara of Italy (INFN) and the Conselho 
Nacional de Pesquisa and Fundacao de Amparo a Pesquisa, Brazil.

\begin{table}
\caption{90\% C.L. Upper Limits for Positively Charged Strangelets}
\begin{tabular}{|c|cc|c|}
\hline
Charge (Z) & Mass No. (A) && 90\% C.L. Upper Limit \\ \hline
+1   &   6    &&   $1.8 \times 10^{-8}$   \\
+1   &  10    &&   $1.2 \times 10^{-8}$   \\
+1   &  20    &&   $7.2 \times 10^{-9}$   \\
+1   &  40    &&   $6.3 \times 10^{-9}$   \\
+1   & 100    &&   $7.0 \times 10^{-9}$   \\ \hline
+2   &   6    &&   $1.5 \times 10^{-8}$   \\
+2   &  10    &&   $9.8 \times 10^{-9}$   \\
+2   &  20    &&   $7.9 \times 10^{-9}$   \\
+2   &  40    &&   $7.4 \times 10^{-9}$   \\
+2   & 100    &&   $7.7 \times 10^{-9}$   \\ \hline
+3   &   6    &&   $1.7 \times 10^{-8}$   \\
+3   &  10    &&   $1.1 \times 10^{-8}$   \\
+3   &  20    &&   $8.8 \times 10^{-9}$   \\
+3   &  40    &&   $8.6 \times 10^{-9}$   \\
+3   & 100    &&   $9.5 \times 10^{-9}$   \\ \hline\hline
\end{tabular}
\label{tab:pos_limits}
\end{table}

\begin{table}
\caption{90\% C.L. Upper Limits for Negatively Charged Strangelets}
\begin{tabular}{|c|c|c|c|}
\hline
Charge (Z) & Mass No. (A) & 90\% C.L. Upper Limit ($B\leq0$) & 90\% C.L.
Upper Limit (all B)  \\ \hline
-1   &   5    &   $1.5 \times 10^{-8}$ & $1.0 \times 10^{-8}$ \\
-1   &  20    &   $8.3 \times 10^{-9}$ & $3.4 \times 10^{-9}$ \\
-1   & 100    &   $9.3 \times 10^{-9}$ & $2.9 \times 10^{-9}$ \\ \hline
-2   &   5    &   $6.7 \times 10^{-9}$ & $5.3 \times 10^{-9}$  \\
-2   &   8    &   $5.1 \times 10^{-9}$ & $3.4 \times 10^{-9}$  \\
-2   &  20    &   $3.5 \times 10^{-9}$ & $1.8 \times 10^{-8}$  \\
-2   & 100    &   $3.8 \times 10^{-9}$ & $1.5 \times 10^{-9}$ \\ \hline
-3   &  10    &                        & $7.8 \times 10^{-8}$ \\
-3   &  20    &                        & $1.3 \times 10^{-8}$ \\
-3   & 100    &                  & $4.3 \times 10^{-9}$ \\ \hline\hline
\end{tabular}
\label{tab:neg_limits}
\end{table}

\begin{table}
\caption{90\% C.L. Upper Limits for Neutral Strangelets.}
\begin{tabular}{|c|c|}
\hline
 Mass No. (A) & 90\% C.L. Upper Limit \\ \hline
  6     &   $3.2 \times 10^{-5}$   \\
  8     &   $5.0 \times 10^{-6}$   \\
 10     &   $9.5 \times 10^{-7}$   \\
 15     &   $5.1 \times 10^{-8}$   \\
 20     &   $7.0 \times 10^{-9}$   \\
 40     &   $3.0 \times 10^{-9}$   \\
 60     &   $2.9 \times 10^{-9}$   \\ 
 80     &   $3.1 \times 10^{-9}$   \\ \hline\hline
\end{tabular}
\label{tab:neut_limits}
\end{table}

\end{document}